\documentclass[a4paper,11pt]{article}
\pdfoutput=1 

\usepackage{jheppub} 
\usepackage{amsmath,amssymb}
\usepackage{slashed}
\usepackage{mathtools}
\usepackage{amsfonts}
\usepackage{booktabs, multirow, array}
\usepackage{hhline}
\usepackage{subcaption}
\usepackage{xcolor}
\usepackage{anyfontsize}
\usepackage{comment}
\usepackage[T1]{fontenc} 

\title{\boldmath Fermionic Spectral Functions in a Two-Current Gubser--Rocha Model with Axion Momentum Relaxation}

\author[a]{Cai-E Liu}
\author[a,1]{and Defu Hou}
\note{Corresponding author.}
\author{}
\affiliation[a]{Institute of Particle Physics and Key Laboratory of Quark and Lepton Physics (MOS),
 Central China Normal University, Wuhan 430079, China}

\emailAdd{liucaie88@mails.ccnu.edu.cn}
\emailAdd{houdf@mail.ccnu.edu.cn}

\abstract{
We investigate fermionic spectral functions in a two-current Gubser-Rocha background with linear-axion momentum relaxation. 
Serving as a comparison channel, the dipole coupling $p_1$ is found to produce a soft asymmetric reconstruction without opening a clean hard Mott gap. 
Our main analysis then focuses on two axion-dependent non-minimal couplings modulated by current mixing. 
The $p_2$ tensor channel suppresses the zero-energy Fermi peak and shifts the dominant spectral weight to finite frequencies, resulting in a flat-bottom reconstruction. 
In contrast, the $p_3$ vector channel creates a strong spectral asymmetry in momentum space, generating open, Fermi-arc-like segments. 
Varying the momentum-relaxation strength $\beta/\mu$ and the doping
parameter $\chi$ reveals a channel-dependent response to the two-current
background. Increasing $\beta/\mu$ generally broadens the $p_2$ response,
whereas the one-sided $p_3$ feature varies non-monotonically under changes
of the background parameters. 
These results clarify how current mixing, axion-induced momentum relaxation, and non-minimal fermion couplings jointly reshape holographic fermionic excitations.}

\begin{document} 
\maketitle
\flushbottom

\section{Introduction}
\noindent
The Landau Fermi liquid theory stands as a cornerstone of modern condensed matter physics, providing a successful quasiparticle framework for describing interacting fermionic systems at low temperatures~\cite{landau1957theory,nozieres1963theory,baym2008landau}. 
Its central tenet is that the low-energy excitations of a metal can be described as long-lived quasiparticles adiabatically connected to those of a free Fermi gas. 
However, the discovery of high-temperature copper-oxide superconductors revealed a class of strongly correlated materials whose behavior deviates from this established paradigm~\cite{Bednorz:1986tc,basov2011electrodynamics,Norman_1998}. 
Related strongly correlated systems, including heavy-fermion compounds and certain iron-based superconductors, also exhibit non-Fermi-liquid (NFL) phenomenology. 
Key signatures include linear-in-temperature resistivity in the strange-metal regime and a reconstruction of the low-energy spectral weight in the pseudogap regime, where closed Fermi contours may appear as disconnected Fermi-arc-like segments~\cite{Yoshida:2005zpb,Cremonini:2018xgj,Seo:2018hrc,2019PhRvB..99p1113W,Chakrabarti:2019gow}. 
These observations pose a fundamental theoretical challenge: how can one describe strongly correlated quantum matter when the quasiparticle picture underlying conventional Fermi liquid theory breaks down?

To address the non-perturbative nature of strong correlations, the Anti-de Sitter\slash  Conformal Field Theory (AdS/CFT) correspondence, or holographic duality, has been widely used as a theoretical framework for studying strongly coupled quantum matter~\cite{Maldacena:1997re,Witten:1998qj,Hartnoll:2009sz,Herzog:2009xv}. 
This approach maps certain strongly coupled quantum field theories to classical gravitational systems in higher-dimensional spacetimes. 
A key development in this direction was the introduction of probe fermions in holographic backgrounds, which allows one to compute boundary fermionic Green's functions and the associated spectral function $A(\omega,\mathbf{k})$~\cite{Liu:2009dm,Faulkner:2009wj,Iqbal:2009fd}. 
This provides a useful theoretical probe of non-Fermi-liquid behavior, since $A(\omega,\mathbf{k})$ is closely related to the momentum-resolved single-particle spectral weight measured in angle-resolved photoemission spectroscopy (ARPES)~\cite{2019NatRP...1..609L,Smit:2021dwh,2003RvMP...75..473D,2011PhRvL.106l7005K}.

Subsequent holographic studies have further enriched the spectral phenomenology of strongly coupled fermions. 
Early studies based on translationally invariant Reissner-Nord\-ström Anti-de Sitter (RN-AdS) black holes related non-Fermi-liquid behavior to the near-horizon $AdS_2$ geometry, while momentum conservation led to an infinite DC conductivity. 
This motivated the introduction of momentum relaxation mechanisms, such as holographic lattices~\cite{Donos:2012js,Ling:2013aya,Donos:2014oha,Donos:2013eha,Ling:2014bda,Ling:2016ewj,Cremonini:2019fzz,Balm:2019dxk,Iliasov:2019pav,Hercek:2022wyu} or linear axion fields~\cite{Andrade:2013gsa,Gouteraux:2014hca,Taylor:2014tka,Kim:2015wba,Baggioli:2021xuv,Jeong:2019zab}, which break translational symmetry and lead to finite DC transport. 

Non-minimal couplings of probe fermions have also been used to obtain richer spectral structures. 
In particular, dipole couplings were shown to produce Mott-like spectral gaps~\cite{Edalati:2010ww,Edalati:2010ge,Guarrera:2011my}, a behavior closely related to the appearance of zeros in the fermionic Green's function~\cite{Alsup:2014uca,Ling:2016ewj,Ceplak:2019ymw,Balm:2019dxk}. 
Related symmetry-breaking couplings have been introduced to describe anisotropic spectral reconstruction and Fermi-arc-like features~\cite{Vanacore:2015poa,Oh:2021xbe}. 
Another route is provided by holographic mean-field theory (HMFT)~\cite{Sukrakarn:2023ncp,Yuk:2022lof}, where a bulk bosonic field dual to a fermionic bilinear operator $\langle\bar{\psi}^{(1)}\Gamma\psi^{(2)}\rangle$ condenses and acts as an order parameter~\cite{Byun:2025tnr,Oh:2020cym,Byun:2025ibg,Ghorai:2024nxs}. 
By classifying the tensorial structure of such order parameters, HMFT has been used to realize spectral gaps, pseudogaps, and Fermi-arc-like structures~\cite{Chakrabarti:2021qie,Baishya:2023xbj,Oh:2021xbe}.

Despite these advances, the role of the background geometry in fermionic spectral reconstruction remains an important issue. 
Many holographic studies of fermionic spectral functions are based on RN-AdS-type backgrounds, where non-Fermi-liquid behavior is often tied to the near-horizon $AdS_2$ region. 
The Gubser-Rocha (GR) model provides a useful holographic background for studying strange-metal transport, especially when momentum relaxation is included to obtain linear-in-temperature resistivity~\cite{Gubser:2009qt}. 
Fermionic excitations in GR-type backgrounds have also been shown to exhibit scaling behavior distinct from both conventional Fermi liquids and the  RN-AdS case~\cite{Lu:2024qxj,Gubser:2012yb,Wu:2011cy,Wen:2012ur,Wen:2013zqa,Li:2011sh}. 
Thus, the GR background offers a useful setting for studying fermionic spectral properties beyond the conventional Landau quasiparticle picture.

However, fermionic spectral reconstruction induced by non-minimal couplings has been less explored in two-current GR backgrounds with axion momentum relaxation. 
In such a background, the fermionic response is affected not only by the non-minimal coupling itself, but also by the charge-sector mixing and the dissipative parameters of the geometry. 
It is therefore natural to ask how current-mixing-modulated fermion couplings reshape the spectral function, and how the resulting spectral features are tuned by momentum relaxation and doping.

In this work, we investigate fermionic spectral functions in an extended two-current Einstein-Maxwell-Gubser-Rocha (EMGR) background~\cite{Zhao:2023qms}. 
The background contains linear axion fields, which implement momentum relaxation, and two independent $U(1)$ gauge fields, which allow one to introduce a controllable doping parameter~\cite{Huang:2020iyw,Zhang:2022hxl}. 
On this dissipative two-current background, we first use the dipole coupling $p_1$ as a comparison channel.
We then focus on two axion-related non-minimal couplings whose effective strengths are modulated by the current-mixing factor $F_{\alpha\beta}Y^{\alpha\beta}$.

The tensor channel controlled by \(p_2\) suppresses the zero-energy Fermi peak and transfers the dominant spectral response to finite frequency, producing a flat-bottom spectral reconstruction.
The vector channel controlled by $p_3$ produces a pronounced imbalance of spectral weight between the positive- and negative-momentum branches and gives rise to Fermi-arc-like open spectral segments. 
We further examine how these reconstructed spectra respond to the momentum-relaxation strength $\beta/\mu$ and the doping parameter $\chi$. 
This allows us to characterize how the dissipative two-current background tunes the spectral features induced by the non-minimal fermion couplings. Therefore, this work provides a spectral characterization of how current mixing, axion-induced momentum relaxation, and non-minimal fermion couplings jointly reshape holographic fermionic excitations.

This paper is organized as follows. 
Sec.~\ref{background geometry} introduces the two-current Gubser-Rocha background with linear-axion momentum relaxation. 
Sec.~\ref{Fermionic_Action} presents the probe fermion action and derives the corresponding Dirac equations and retarded Green's function. 
Sec.~\ref{sec:numerical_calculation} discusses the main spectral results, including the dipole comparison response, the $p_2$-induced flat-bottom finite-frequency reconstruction, the $p_3$-induced asymmetric spectral redistribution, and their dependence on $\beta/\mu$ and $\chi$. 
Sec.~\ref{summary} summarizes the results and gives an outlook.

\section{Background Geometry}
\label{background geometry}
Motivated by the Fermi-arc phenomenology discussed in the introduction,
we employ a holographic model to study fermionic spectral functions. The setup uses the two-current Einstein-Maxwell-Gubser-Rocha (EMGR) background, which was also used in Ref.~\cite{Zhao:2023qms} to construct a holographic superconductor. Here we omit the superconducting sector and study the electronic spectra of the dual normal metal. We work with a $(3+1)$-dimensional bulk dual to a $(2+1)$-dimensional boundary theory with homogeneous momentum relaxation. The action reads:
\begin{eqnarray}\label{eq1}
    S=\int d^4x \sqrt{-g}&[R+\frac{6}{L^2}\cosh{\varphi}-\frac{\alpha}{4}e^{\varphi}F_{\mu\nu}F^{\mu\nu}-\frac{\gamma}{4}e^{\varphi}Y_{\mu\nu}Y^{\mu\nu}\\
    \nonumber
    &-\frac{3}{2}(\partial \varphi)^2-\sum_{I=x,y} (\partial \psi_I)^2]
\end{eqnarray}
where $R$ is the Ricci scalar, and $L$ is the AdS radius, which will be set to unity later.
The field strengths are defined as $F=dA$ and $Y=dB$, where $d$ denotes the exterior derivative acting on the vector potential one-forms $A$ and $B$, respectively.
The scalar field $\varphi$ serves as a dilaton that ensures the entropy is proportional to the temperature in the near-zero-temperature limit.
$F_{\mu\nu}$ and $Y_{\mu\nu}$ denote the field strengths associated with $A_\mu$ and $B_\mu$, respectively, which are introduced to provide finite chemical potentials.
In addition, two axion fields $\psi_I$ ($\psi_x=\beta x,\,\psi_y=\beta y$) are included to break translational invariance.
From the action \eqref{eq1}, we obtain the following equations of motion:
\begin{align}
    &R_{\mu\nu}-\frac{1}{2}R g_{\mu\nu}-\frac{3g_{\mu\nu}}{L^2}\cosh{\varphi}=\frac{\alpha}{2}e^{\varphi}T_{\mu\nu}^{(A)}+\frac{\gamma}{2}e^\varphi T_{\mu\nu}^{(B)}+T_{\mu\nu}^{(\varphi)}+T_{\mu\nu}^{(\psi)}\\
&\nabla_\mu \nabla^\mu \varphi
+\frac{2}{L^2}\sinh{\varphi}-\frac{1}{12}e^\varphi (\alpha F_{\mu\nu}F^{\mu\nu}+\gamma Y_{\mu\nu}Y^{\mu\nu})=0\\
&\nabla_\mu(\alpha e^\varphi F^{\mu\nu})=0\\
&\nabla_\mu(\gamma e^\varphi Y^{\mu\nu})=0\\
&\nabla_\mu \nabla^\mu \psi_I=0
\end{align}
The explicit forms of the energy–momentum tensors 
$T_{\mu\nu}^{(A)}$, $T_{\mu\nu}^{(B)}$, 
$T_{\mu\nu}^{(\varphi)}$, and $T_{\mu\nu}^{(\psi)}$ are given by:
\begin{align}
T_{\mu\nu}^{(A)}&=F_{\mu\rho}F_{\nu}^{~\rho}-\frac{1}{4}g_{\mu\nu}F^2\\
T_{\mu\nu}^{(B)}&=Y_{\mu\rho}Y_{\nu}^{~\rho}-\frac{1}{4}g_{\mu\nu}Y^2\\
T_{\mu\nu}^{(\varphi)}&=\frac{3}{2}\partial_{\mu} \varphi \partial_{\nu}\varphi-\frac{3}{4}(\partial \varphi)^2 g_{\mu\nu}\\
T_{\mu\nu}^{(\psi)}&=\sum_{I=x,y}[(\partial_{\mu}\psi_I \partial_{\nu}\psi_I)-\frac{1}{2}g_{\mu\nu}(\partial \psi_I)^2]
\end{align}
We assume the following ansatz:
\begin{align}
&\mathrm{d}s^2 = -g_{tt}\mathrm{d}t^2 + g_{rr}\mathrm{d}r^2 + g_{xx}\mathrm{d}x^2 + g_{yy}\mathrm{d}y^2 \\
&= -\frac{\tilde{r}^2g(\tilde{r})h(\tilde{r})}{L^2}\,\mathrm{d}\tilde{t}^2 + \frac{L^2}{\tilde{r}^2g(\tilde{r})h(\tilde{r})}\,\mathrm{d}\tilde{r}^2 + \frac{\tilde{r}^2g(\tilde{r})}{L^2}\,\mathrm{d}\tilde{x}^2 + \frac{\tilde{r}^2g(\tilde{r})}{L^2}\,\mathrm{d}\tilde{y}^2\,,\\
&A = A_t(\tilde{r})d\tilde{t}, \quad B = B_t(\tilde{r})d\tilde{t}, \quad \psi_x = \tilde{\beta}\tilde{x}, \quad \psi_y = \tilde{\beta}\tilde{y}, \quad \varphi = \varphi(\tilde{r})
\end{align}
 Our model admits the following solution,
 \begin{align}\label{solution}
    &\varphi=\frac{1}{3}\log{g(\tilde{r})}\qquad g(\tilde{r})=(1+\frac{1}{\tilde{r}})^{\frac{3}{2}}\\ \nonumber
   & A_t(\tilde{r})=\tilde{\mu}(1-\frac{1+\tilde{r_h}}{1+\tilde{r}})\qquad B_t(\tilde{r})=\tilde{\mu}_B(1-\frac{1+\tilde{r_h}}{1+\tilde{r}})\\ \nonumber
   &\tilde{\mu}=\frac{\sqrt{3}\sqrt{1-\tilde{\beta}^2L^4+2\tilde{r_h}+{\tilde{r_h}}^2}}{\sqrt{L^2\alpha+L^2 \tilde{r_h}\alpha+L^2\chi^2\gamma+L^2\tilde{r_h}\chi^2\gamma}}\\ \nonumber
   &\tilde{\mu}_B=\chi \frac{\sqrt{3}\sqrt{1-\tilde{\beta}^2L^4+2\tilde{r_h}+{\tilde{r_h}}^2}}{\sqrt{L^2\alpha+L^2 \tilde{r_h}\alpha+L^2\chi^2\gamma+L^2\tilde{r_h}\chi^2\gamma}}\\ \nonumber
   & h(\tilde{r})=1-\frac{(1+\tilde{r_h})^3}{(1+\tilde{r})^3}-\frac{\tilde{\beta}^2 L^4}{(1+\tilde{r})^2}\left(1-\frac{1+\tilde{r_h}}{1+\tilde{r}}\right)   
\end{align}
Here, $r_h$ denotes the horizon radius, $\beta$ represents the strength of momentum relaxation, and $Q$ is a physical parameter related to the charge carrier densities dual to the gauge fields. Specifically, the charge carrier densities corresponding to $A_\mu$ and $B_\mu$ are given by $\rho_A = \mu (Q + r_h)$ and $\rho_B = \mu_B (Q + r_h)$, respectively. In the present solution, the parameter $\chi$ controls the relative weight
of the two charge sectors. In particular, one has
$\mu_B/\mu=\rho_B/\rho_A=\chi$, so that $\chi$ serves as the doping
parameter throughout the following analysis. After introducing the following dimensionless variables
\[
\tilde{r} := \frac{r}{Q}, \quad 
\tilde{t} := t Q, \quad 
\tilde{x} := x Q, \quad 
\tilde{y} := y Q, \quad 
\tilde{\beta} := \frac{\beta}{Q},\quad \tilde{\mu}:=\frac{\mu}{Q}\quad
\tilde{\mu_B}:=\frac{\mu_B}{Q}
\]
the rescaled horizon radius $\tilde{r}_h$ satisfies $h(\tilde{r}_h) = 0$, and $\tilde{\beta}$ characterizes the strength of momentum relaxation or translational symmetry breaking. For simplicity, we set $L=1$ and $\alpha=\gamma=1$ in the following discussion.

The temperature is obtained from  \eqref{solution} as
\begin{align}
f(\tilde{r})&=\tilde{r}^2g(\tilde{r})h(\tilde{r})\\
T&=\frac{|f'(\tilde{r_h})|}{4\pi}=Q\frac{\sqrt{\tilde{r_h}}(3(1+\tilde{r_h})^2-\tilde{\beta}^2)}{4\pi(1+\tilde{r_h})^{\frac{3}{2}}}=:Q\tilde{T}
\end{align}

In this paper, we explore the spectral function at fixed chemical potential. For this purpose we define:
\begin{align}
&\bar{T}:=\frac{T}{\mu}=\frac{\tilde{T}}{\tilde{\mu}}=\frac{\sqrt{\tilde{r_h}(1+\chi^2)}(3(1+\tilde{r_h})^2-\tilde{\beta}^2)}{4\pi(1+\tilde{r_h})\sqrt{3(1+\tilde{r_h})^2-3\tilde{\beta}^2}}\\
&\bar{\beta}:=\frac{\beta}{\mu}=\frac{\tilde{\beta}}{\tilde{\mu}}=\frac{\tilde{\beta}\sqrt{1+\chi^2}\sqrt{1+\tilde{r_h}}}{\sqrt{3(1+\tilde{r_h})^2-3\tilde{\beta}^2}}
\end{align}

\section{Fermionic Action and Interaction Terms}
\label{Fermionic_Action}
To study the fermionic spectral function, we consider a probe fermion propagating in the gravitational background defined in the previous section. We adopt a unified framework where the bulk fermionic dynamics are generally described by the action:
\begin{equation}
\label{eq:general_action}
S_{\mathrm{bulk}} = \int \mathrm{d}^{4}x \ \sqrt{-g} \,i \overline{\psi}\,\Big( \Gamma^{a}D_{a} - m_\psi +\mathcal{L}_{\text{int}} \Big)\psi \,.
\end{equation}
Here, $e_{\mu}^{a}$ and $\omega_{\mu\nu}^{~~a}$ denote the vielbein and the spin connection one-forms, respectively. $A_\mu$ is the bulk electromagnetic potential, and $q$ and $m_\psi$ correspond to the charge and mass of the bulk fermion. In our conventions, $a,b$ are usual spacetime abstract index and $\mu,\nu$ are the tangent-space index.

Before proceeding with the specific interaction scenarios, we note the definitions for the covariant derivative $D_a$ and the Maxwell tensor $\slashed{F}$, along with the essential gamma matrix identities:
\begin{align}
&D_a = \partial_a + \frac{1}{4} (\omega_{\mu\nu})_a \Gamma^{\mu\nu} - iqA_a \,, \\
&(\omega_{\mu\nu})_a=(e_\mu)^b\nabla_a (e_\nu)_b\,, \qquad \slashed{F}=\frac{1}{2}\Gamma^{\mu\nu}(e_{\mu})^{a}(e_{\nu})^{b}F_{ab} \,, \\
&\Gamma^{\mu\nu} = \frac{1}{2}[\Gamma^\mu, \Gamma^\nu ]\,, \qquad \{\Gamma^\mu, \Gamma^\nu\} = 2 \eta_{\mu\nu} \,, \\
& \vec{\Gamma} = (\Gamma^x, \Gamma^y)\,, \qquad \Gamma = \Gamma^r \Gamma^t(\hat{n} \cdot \vec{\Gamma})=\Gamma^r\Gamma^t\Gamma^x \,.
\end{align}
The interaction term $\mathcal{L}_{\text{int}}$ captures the coupling between the fermion and the bosonic background. We specify below the dipole comparison channel and the two axion-related
interaction channels considered in this work.

\subsection{Dipole Coupling}
\label{subsec:DP_coupling}
Before introducing the axion-induced interaction channels, we first consider the  dipole coupling as a comparison channel. 
The Pauli-type term $i p_1 \slashed{F}$ has been widely used in holographic fermion systems to study dipole-induced spectral reconstruction and the interplay between poles and zeros of the fermionic Green's function. 
The corresponding interaction term is
\begin{equation}
\mathcal{L}_{\text{int}}^{(\mathrm{I})} =- i p_{1}\,\slashed{F} .
\end{equation}
From the above action, we can derive the following Dirac equation:
\begin{equation}
    \left(\Gamma^a D_a - m_{\psi} - ip_1 \slashed{F}\right) \psi = 0 \,.
\end{equation}
Expanding this explicitly, one has:
\begin{align}
   &(\Gamma^t\sqrt{g^{tt}}(\partial_t-iq A_t)+\Gamma^r\sqrt{g^{rr}}\partial_r+\Gamma^x\sqrt{g^{xx}}\partial_x+\Gamma^y\sqrt{g^{yy}}\partial_y)\psi \nonumber \\
   &\qquad+\sqrt{g^{rr}}\partial_r\ln{(-g_{tt}g_{ii}})^{\frac{1}{4}}\Gamma^r\psi
   -ip_1\frac{\Gamma^{rt}A_t'}{\sqrt{g_{tt}g_{rr}}}\psi-m_\psi \psi=0 \,.
\end{align}
We then perform the Fourier transformation $\psi=(-gg^{rr})^{-\frac{1}{4}}\Phi$,
\begin{align}
\Phi(\tilde{r}, t, x, y) = \int \frac{d\omega dk_x dk_y}{(2\pi)^3} e^{-i\omega t + i k_x x + i k_y y} \phi(\tilde{r}, \vec{k}).
\end{align}
By choosing $\hat{n}=\hat{x}$ and using the following form of the gamma matrices:
\begin{align}
\Gamma^{\underline{r}} &= 
\left( \begin{array}{cc} 
-\sigma^3 & 0 \\ 
0 & -\sigma^3 
\end{array} \right), &
\Gamma^{\underline{t}} &= 
\left( \begin{array}{cc} 
i\sigma^1 & 0 \\ 
0 & i\sigma^1 
\end{array} \right), \\
\Gamma^{\underline{x}} &= 
\left( \begin{array}{cc} 
-\sigma^2 & 0 \\ 
0 & \sigma^2 
\end{array} \right), &
\Gamma^{\underline{y}} &= 
\left( \begin{array}{cc} 
0 & \sigma^2 \\ 
\sigma^2 & 0 
\end{array} \right),\qquad
\Gamma^{\underline{5}}= 
\left( \begin{array}{cc} 
0 & i \sigma_2 \\ 
-i\sigma_2 & 0 
\end{array} \right),
\end{align}
and splitting the 4-component spinor into two 2-component spinors as $\phi=(\phi_1,\phi_2)^T$, we have the following radial equations:
\begin{align}
&\frac{\sqrt{g_{ii}}}{\sqrt{g_{rr}}}\partial_r\otimes \begin{pmatrix}\phi_1 \\ \phi_2 \end{pmatrix}-i\frac{\sqrt{g_{ii}}}{\sqrt{g_{tt}}}(\omega + qA_t) \sigma_2 \otimes\begin{pmatrix} \phi_1 \\ \phi_2 \end{pmatrix}+m_\psi\sqrt{g_{ii}}\sigma_3 \otimes\begin{pmatrix} \phi_1 \\ \phi_2 \end{pmatrix}\pm k_x\sigma_1\otimes \begin{pmatrix} \phi_1 \\ \phi_2 \end{pmatrix} \nonumber \\
&\qquad +\frac{p_1A_t'\sqrt{g_{ii}}}{\sqrt{g_{tt}g_{rr}}}\sigma_1\otimes \begin{pmatrix} \phi_1 \\ \phi_2 \end{pmatrix}-k_y\sigma_1 \begin{pmatrix} \phi_2 \\ \phi_1 \end{pmatrix}=0 \,.
\end{align}
In these equations, the symbol $\otimes$ serves as a shorthand for the element-wise action of the Pauli matrices. For instance, the expression $\sigma_1 \otimes (\phi_2, \phi_1)^T$ is equivalent to the column vector $(\sigma_1 \phi_2, \sigma_1 \phi_1)^T$, signifying that the operator $\sigma_1$ acts uniformly on the internal degrees of freedom of both $\phi_1$ and $\phi_2$.

Furthermore, by the decomposition $\phi_I=(y_I,z_I)^T$ with $I=1,2$, the above Dirac equation can be expressed as:
\begin{align}
&\frac{\sqrt{g_{ii}} }{\sqrt{g_{rr}}}\otimes\begin{pmatrix} y_1' \\ z_1' \end{pmatrix} \pm m_\psi \sqrt{g_{ii}}\otimes \begin{pmatrix} y_1 \\ z_1 \end{pmatrix}\mp\frac{\sqrt{g_{ii}}}{\sqrt{g_{tt}}}(\omega+q A_t)\otimes \begin{pmatrix} z_1 \\ y_1 \end{pmatrix}+ k_x\begin{pmatrix} z_1 \\ y_1 \end{pmatrix} \nonumber \\
&\qquad +\frac{p_1  \sqrt{g_{ii}}  A_t'}{\sqrt{g_{rr} g_{tt}}}\otimes\begin{pmatrix} z_1 \\ y_1 \end{pmatrix}-k_y\otimes \begin{pmatrix} z_2 \\ y_2 \end{pmatrix}=0 \,, \\
&\frac{\sqrt{g_{ii}} }{\sqrt{g_{rr}}}\otimes\begin{pmatrix} y_2' \\ z_2' \end{pmatrix} \pm m_\psi \sqrt{g_{ii}}\otimes \begin{pmatrix} y_2 \\ z_2 \end{pmatrix}\mp\frac{\sqrt{g_{ii}}}{\sqrt{g_{tt}}}(\omega+q A_t)\otimes \begin{pmatrix} z_2 \\ y_2 \end{pmatrix}- k_x\begin{pmatrix} z_2 \\ y_2 \end{pmatrix} \nonumber \\
&\qquad +\frac{p_1  \sqrt{g_{ii}}  A_t'}{\sqrt{g_{rr} g_{tt}}}\otimes\begin{pmatrix} z_2 \\ y_2 \end{pmatrix}-k_y\otimes \begin{pmatrix} z_1 \\ y_1 \end{pmatrix}=0 \,.
\end{align}

At the horizon, the leading term of the equation of motion is given by:
\begin{align}
\partial_r \begin{pmatrix} y_I(r,\vec{k}) \\ z_I(r,\vec{k}) \end{pmatrix} + \frac{\omega}{4\pi T} \frac{1}{(r-r_h)} \begin{pmatrix} z_I(r,\vec{k}) \\ y_I(r,\vec{k}) \end{pmatrix} = 0 \,.
\end{align}

To obtain the retarded Green's function at the AdS boundary, we impose
the infalling boundary condition at the horizon, 
\begin{align}
    \begin{pmatrix} y_I(r,\vec{k}) \\ z_I(r,\vec{k}) \end{pmatrix} \sim B_j\begin{pmatrix} 1\\ -i  \end{pmatrix}(r-r_h)^{-\frac{i\omega}{4\pi T}} \,.
\end{align}
Near the AdS boundary, the asymptotic behaviour of the Dirac equation has the following form:
\begin{align}\label{bdc}
\begin{pmatrix} y_I \\ z_I \end{pmatrix} \sim C_j r^{-m_{\psi}} \begin{pmatrix} 1 \\ 0 \end{pmatrix} + D_j r^{m_{\psi}} \begin{pmatrix} 0 \\ 1 \end{pmatrix} \,.
\end{align}
Due to the mixing of spinor components under the considered coupling, extracting the Green's function requires the prescription detailed in Refs.~\cite{Vanacore:2015poa,Guarrera:2011my,Faulkner:2009am}. We integrate two linearly independent infalling
solutions from the horizon and numerically extract the coefficients
$C$ and $D$ from the asymptotic expansion in Eq.~\eqref{bdc}. From the holographic dictionary, the retarded Green function is given by:
\begin{equation}
    G_R=\mathcal{C}\mathcal{D}^{-1} \,,
\end{equation}
where I and II label the two linearly independent infalling solutions. We have defined:
\begin{equation}
    G_{R}= \left( \begin{array}{cc} G_{11} & G_{12} \\ G_{21} & G_{22}\end{array} \right),\qquad \mathcal{C}=\left( \begin{array}{cc} c_1^\mathrm{I} & c_1^\mathrm{II} \\ c_2^\mathrm{I} & c_2^\mathrm{II} \end{array} \right),\qquad
    \mathcal{D}= \left( \begin{array}{cc} d_1^\mathrm{I} & d_1^\mathrm{II} \\ d_2^\mathrm{I} & d_2^\mathrm{II} \end{array} \right) \,.
\end{equation}
For numerical convenience, we can define the following matrices:
\begin{align}
Y = \left( \begin{array}{cc} y_1^\mathrm{I} & y_1^\mathrm{II} \\ y_2^\mathrm{I} & y_2^\mathrm{II} \end{array} \right), \quad
Z = \left( \begin{array}{cc} z_1^\mathrm{I} & z_1^\mathrm{II} \\ z_2^\mathrm{I} & z_2^\mathrm{II} \end{array} \right), \quad
\tilde{G} = Y Z^{-1} \,.
\end{align}
Finally, the flow equations follow:
\begin{align}
    \frac{\tilde{G}_{11}'}{\sqrt{g_{rr}}} + 2m_\psi \tilde{G}_{11} &= -\frac{k_{x}}{\sqrt{g_{xx}}}(1+\tilde{G}_{12}\tilde{G}_{21}-\tilde{G}_{11}^2)-\frac{k_y \tilde{G}_{11}(\tilde{G}_{12}+\tilde{G}_{21})}{\sqrt{g_{yy}}}\nonumber \\ 
    &\quad +\frac{(\omega+q A_t)(1+\tilde{G}_{11}^2+\tilde{G}_{12}\tilde{G}_{21})}{\sqrt{g_{tt}}}+A_t'\frac{(p_1)(-1+\tilde{G}_{11}^2)+(p_1)\tilde{G}_{12}\tilde{G}_{21}}{\sqrt{g_{rr}g_{tt}}} \label{eq:G11} \,, \\ 
    \frac{\tilde{G}_{12}'}{\sqrt{g_{rr}}} + 2m_\psi \tilde{G}_{12} &= \frac{k_{y}}{\sqrt{g_{yy}}}(1-\tilde{G}_{12}^2-\tilde{G}_{11}\tilde{G}_{22})+\frac{k_x \tilde{G}_{12}(\tilde{G}_{11}-\tilde{G}_{22})}{\sqrt{g_{xx}}}\nonumber \\ 
    &\quad +\frac{(\omega+q A_t)(\tilde{G}_{11}\tilde{G}_{12}+\tilde{G}_{12}\tilde{G}_{22})}{\sqrt{g_{tt}}}+A_t'\frac{(p_1)\tilde{G}_{11}\tilde{G}_{12}+(p_1)\tilde{G}_{12}\tilde{G}_{22}}{\sqrt{g_{rr}g_{tt}}} \label{eq:G12} \,, \\
    \frac{\tilde{G}_{21}'}{\sqrt{g_{rr}}} + 2m_\psi \tilde{G}_{21} &= \frac{k_{y}}{\sqrt{g_{yy}}}(1-\tilde{G}_{21}^2-\tilde{G}_{11}\tilde{G}_{22})+\frac{k_x \tilde{G}_{21}(\tilde{G}_{11}-\tilde{G}_{22})}{\sqrt{g_{xx}}}\nonumber \\ 
    &\quad +\frac{(\omega+q A_t)(\tilde{G}_{11}\tilde{G}_{21}+\tilde{G}_{21}\tilde{G}_{22})}{\sqrt{g_{tt}}}+A_t'\frac{(p_1)\tilde{G}_{11}\tilde{G}_{21}+(p_1)\tilde{G}_{21}\tilde{G}_{22}}{\sqrt{g_{rr}g_{tt}}} \label{eq:G21} \,, \\
    \frac{\tilde{G}_{22}'}{\sqrt{g_{rr}}} + 2m_\psi \tilde{G}_{22} &= \frac{k_{x}}{\sqrt{g_{xx}}}(1+\tilde{G}_{12}\tilde{G}_{21}-\tilde{G}_{22}^2)-\frac{k_y \tilde{G}_{22}(\tilde{G}_{12}+\tilde{G}_{21})}{\sqrt{g_{yy}}}\nonumber \\ 
    &\quad +\frac{(\omega+q A_t)(1+\tilde{G}_{22}^2+\tilde{G}_{12}\tilde{G}_{21})}{\sqrt{g_{tt}}}+A_t'\frac{(p_1)(-1+\tilde{G}_{22}^2)+(p_1)\tilde{G}_{12}\tilde{G}_{21}}{\sqrt{g_{rr}g_{tt}}} \label{eq:G22} \,.
\end{align}

Imposing the requirement of regularity on the flow equations at the horizon straightforwardly leads to the boundary conditions
\begin{align}
\tilde{G}^{r \rightarrow r_h}\simeq \left( \begin{array}{cc} i & 0 \\ 0 & i \end{array} \right) \,.
\end{align}
After solving the evolution equations, we can read off the boundary Green's function as
\begin{align}
    G_R = \lim_{r \to \infty} r^{2m_{\psi}} \tilde{G} \,.
\end{align}
The spectral function is defined as
\begin{equation}\label{spec-eq}
A(\omega,k_x,k_y)=\mathrm{Tr}[\mathrm{Im}(G_R(\omega,k_x,k_y))] \,.
\end{equation}
For the dipole channel, along $k_y=0$, Eqs.~\eqref{eq:G11}--\eqref{eq:G22} imply that the off-diagonal components vanish and that the diagonal components are related by momentum reflection,
$G_{11}^{R}(\omega,k_x)=G_{22}^{R}(\omega,-k_x)$.
Accordingly, when tracking the evolution of an individual momentum branch below, we use $\operatorname{Im}G_{22}$ as a representative component, while the total spectral function remains defined by Eq.~\eqref{spec-eq}.

\subsection{Current-mixing-modulated axion-related couplings}
\label{subsec:Interaction_Driven}

The interaction terms  \(L_{\rm int}\) introduced in this subsection are motivated by holographic Ginzburg--Landau--Wilson-type constructions and by recent studies on holographic Mott-like spectral structures~\cite{Oh:2020cym,Ghorai:2024nxs}. 
These works showed that different gamma-matrix structures and non-minimal fermion couplings can lead to distinct spectral responses, including spectral gaps, pseudogaps, flat-band-like features, and Fermi-arc-like structures. 
In the present work, we use these mechanisms as guidance, but embed the corresponding interaction structures in a two-current Gubser-Rocha background with linear-axion momentum relaxation.

A central feature of this construction is that the new interaction channels are modulated by the two-current contraction \(F_{\alpha\beta}Y^{\alpha\beta}\). 
This factor couples the probe fermion sector to the two bulk gauge fields and makes the effective interaction strength sensitive to current mixing in the background. 
In addition, because the couplings involve the linear axion fields,
\[
\psi_I=\beta \delta_{Ii}x^i,
\]
these couplings also depend on the momentum-relaxation sector. 
We consider two such channels: a tensor-type coupling controlled by $p_2$ and a vector/axion-gradient coupling controlled by $p_3$.

The corresponding interaction terms are
\begin{subequations}
\label{eq:axion_interactions}
    \begin{align}
        \mathcal{L}^{\text{(II-a)}}_{\text{int}}  
        &= -i p_2 F_{\alpha\beta} Y^{\alpha\beta} 
        \Gamma^{\mu\nu} H_{\mu\nu}  \,, \label{eq:Lint1} \\
        \mathcal{L}^{\text{(II-b)}}_{\text{int}} 
        &= -i p_3 F_{\alpha\beta} Y^{\alpha\beta} 
       \sum_{I=1}^{2} (\Gamma^{\mu} \partial_{\mu}\psi_I) \Gamma^5 \,. \label{eq:Lint2}
    \end{align}
\end{subequations}
Here the auxiliary tensor is defined by
\begin{equation}
H_{\mu\nu}
=
\frac{1}{2}\epsilon^{IJ}
\partial_{\mu}\psi_I
\partial_{\nu}\psi_J ,
\end{equation}
which is constructed from the axion gradients and encodes the spatial structure associated with momentum relaxation. 
The prefactor $F_{\alpha\beta}Y^{\alpha\beta}$ represents the mixing between the two bulk gauge sectors.

Before analyzing the spectral consequences of these couplings, it is useful to examine their UV behavior. 
Near the asymptotic AdS boundary, the gauge-field derivatives behave as
\[
A_t', B_t'\sim \mathcal{O}(r^{-2}),
\]
while
\[
g_{tt},g_{ii}\sim \mathcal{O}(r^2),
\qquad
g_{rr}\sim \mathcal{O}(r^{-2}) .
\]
Therefore, the effective radial prefactor of the $p_2$ channel scales schematically as
\[
\lambda_2(r)
\sim
\frac{\beta^2 A_t'B_t'}
{g_{xx}g_{rr}g_{tt}}
\sim
\mathcal{O}(r^{-6}) ,
\]
whereas the corresponding prefactor of the $p_3$ channel scales as
\[
\lambda_3(r)
\sim
\frac{\beta A_t'B_t'}
{g_{rr}g_{tt}\sqrt{g_{ii}}}
\sim
\mathcal{O}(r^{-5}) .
\]
Thus both axion-related non-minimal couplings vanish in the UV region. 
They do not modify the leading source--response structure of the bulk spinor, and the retarded Green's function can be extracted using the  holographic prescription. 
Their influence on the spectral function arises from the bulk radial evolution, especially from the interior and near-horizon regions.

We now briefly distinguish the roles of the two interaction channels.
The $p_2$ interaction contains the tensor structure
$\Gamma^{\mu\nu}H_{\mu\nu}$ and is modulated by the current-mixing
factor $F_{\alpha\beta}Y^{\alpha\beta}$. Motivated by related tensorial
couplings discussed in Refs.~\cite{Oh:2020cym,Ghorai:2024nxs}, this
channel provides a mechanism through which the axion sector and current
mixing modify the fermionic spectral response.

By contrast, the $p_3$ interaction contains the axion-gradient structure
$(\Gamma^\mu\partial_\mu\psi_I)\Gamma^5$, corresponding to a
vector/axial-type deformation of the probe-fermion sector. Related
axial-vector couplings have been used to generate anisotropic and
Fermi-arc-like spectral structures in holographic fermion
models~\cite{Oh:2020cym}. Here, this interaction is embedded in the
two-current dissipative background, allowing us to examine how the
axion sector and current mixing modify the momentum-space spectral response.

With the interaction terms specified, the numerical procedure follows the holographic prescription outlined above.
We solve the flow equations detailed in Appendix~\ref{app:flow_equations} with the infalling boundary condition at the horizon,
\[
\tilde{G}(r\to r_h)\simeq i\mathbb{I}_{2\times2}.
\]
The boundary value of the flow matrix determines the retarded Green's function, which will be used below to analyze the interaction-induced spectral reconstructions.

\section{Numerical Results}
\label{sec:numerical_calculation}

\begin{figure}[htb!]
    \centering
    \setlength{\tabcolsep}{4pt} 
        \begin{tabular}{ccc}
        \phantomsubcaption\label{fig:chi0.4}\includegraphics[width=0.32\textwidth]{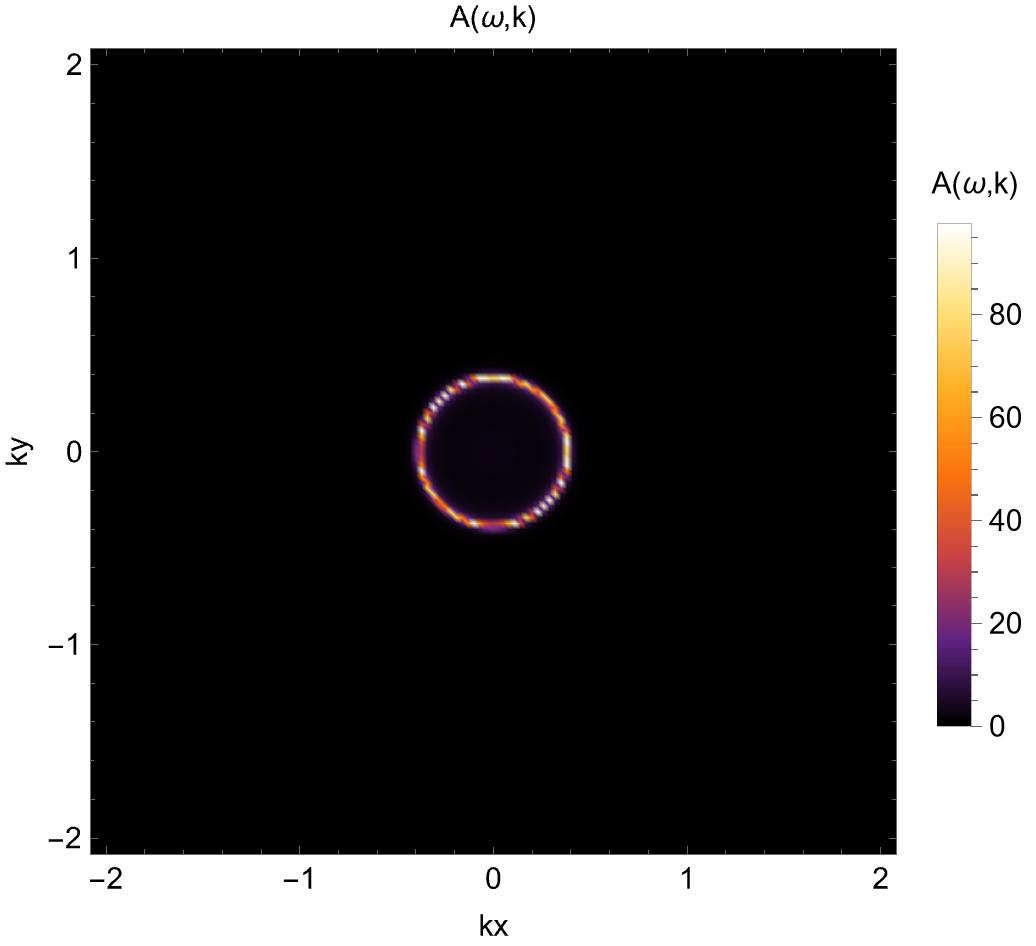} &
        \phantomsubcaption\label{fig:chi1.0}\includegraphics[width=0.32\textwidth]{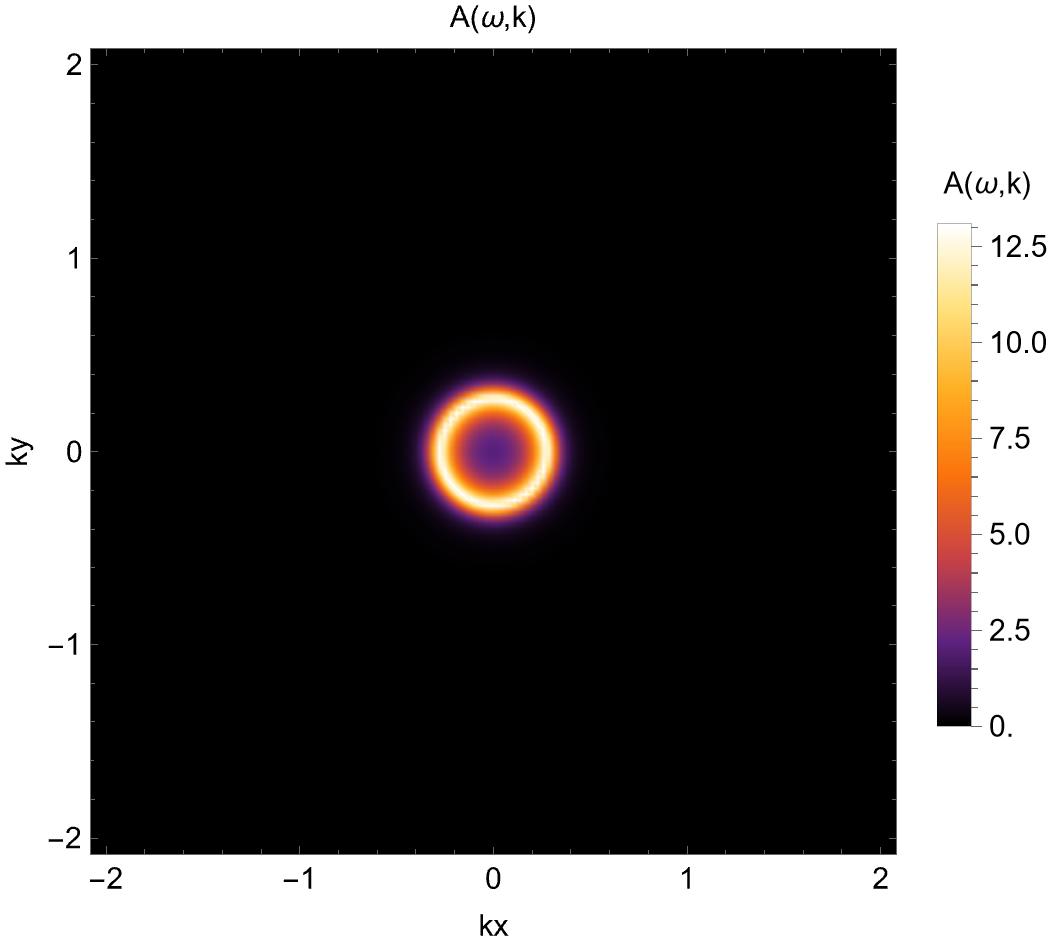} &
        \phantomsubcaption\label{fig:chi2.3}\includegraphics[width=0.32\textwidth]{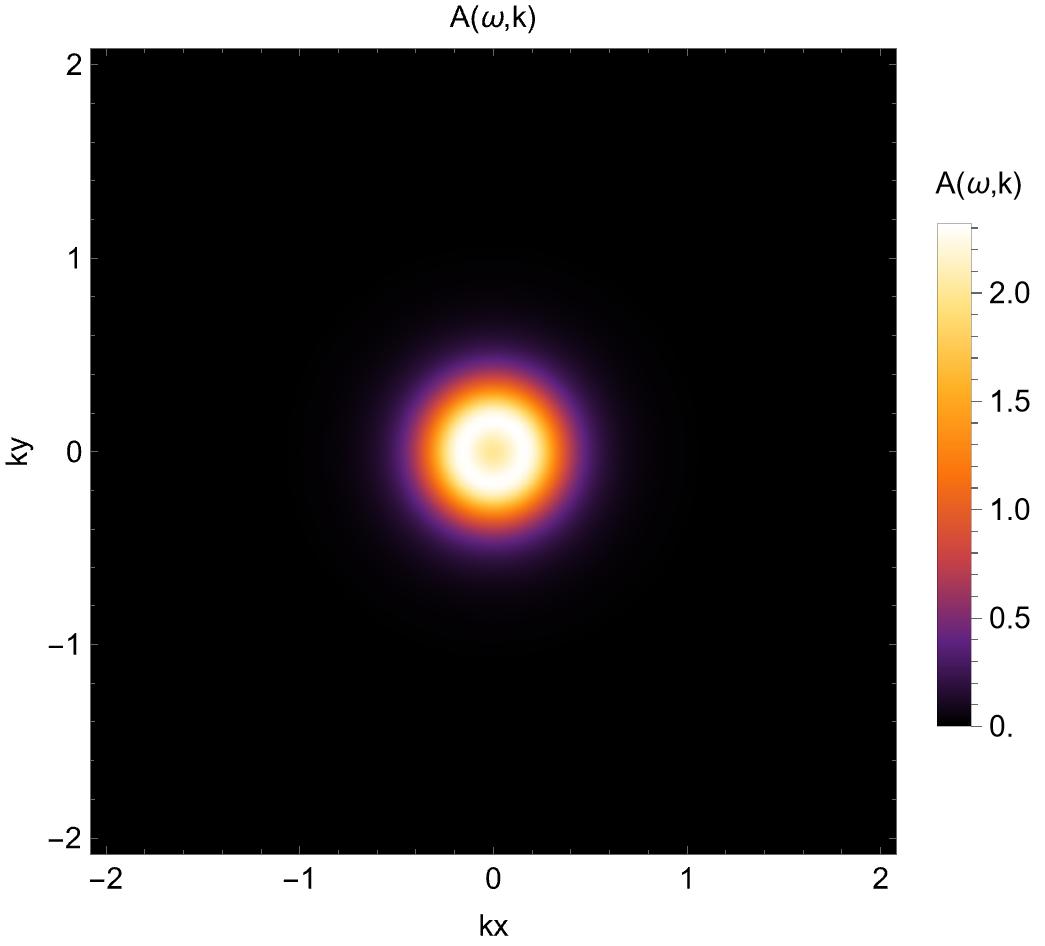}
        \\
        (a) $\chi=0.4$ & (b)$\chi=1.0$&(c)$\chi=2.3$ 
        \\

\phantomsubcaption\label{fig:beta0}\includegraphics[width=0.32\textwidth]{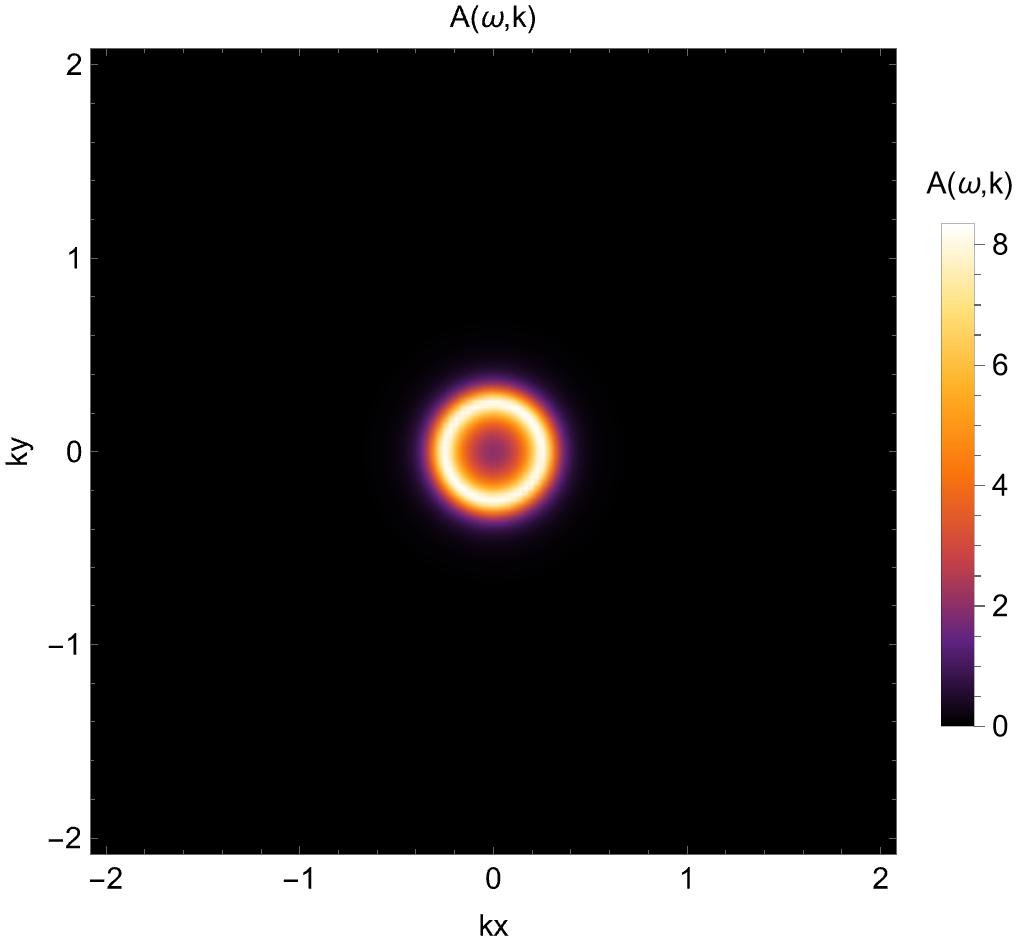} &
        \phantomsubcaption\label{fig:beta4}\includegraphics[width=0.32\textwidth]{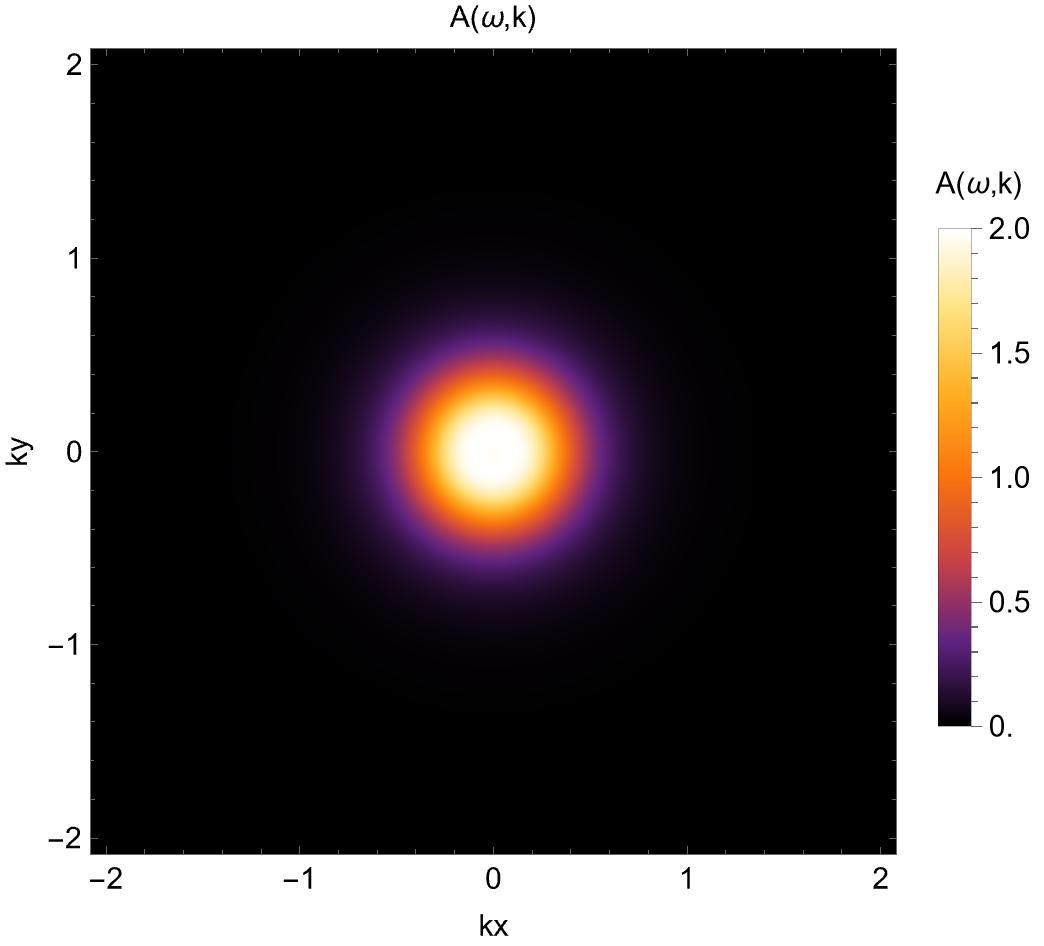} &
        \phantomsubcaption\label{fig:beta8}\includegraphics[width=0.32\textwidth]{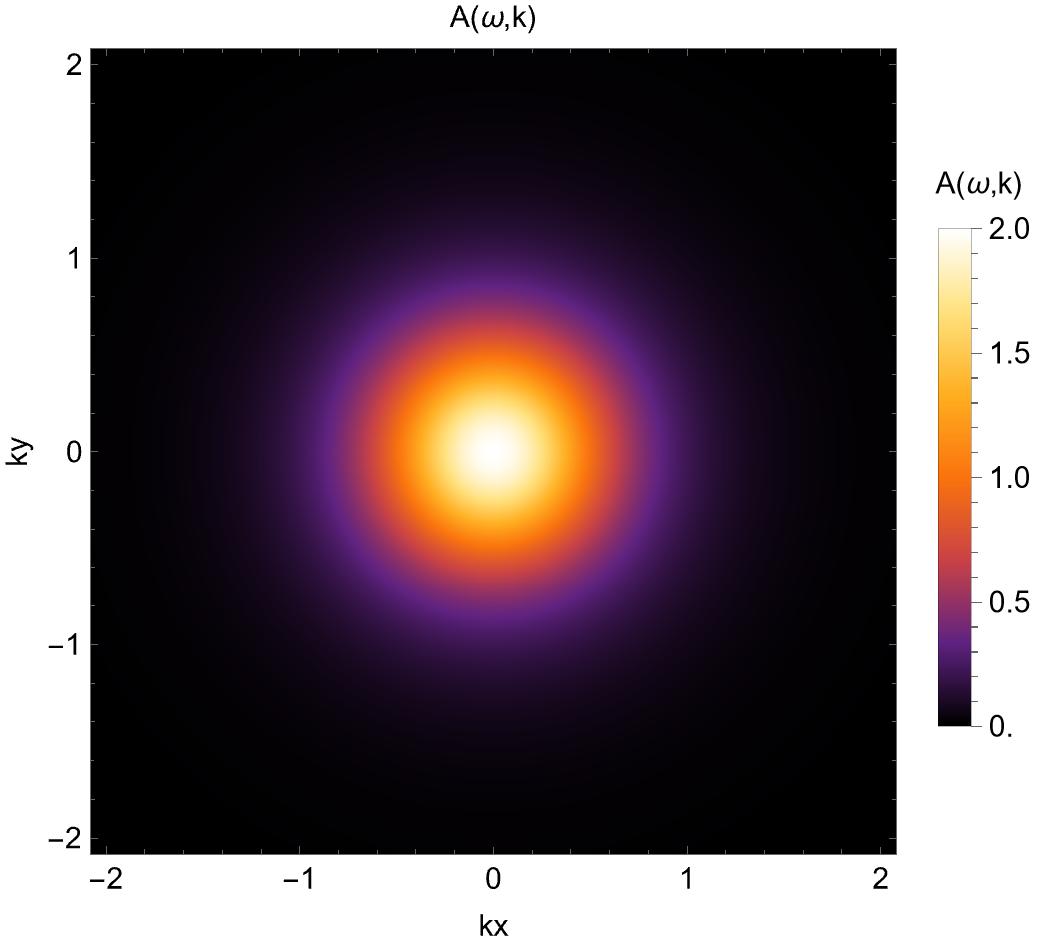}  \\
        (d) $\beta/\mu=0.4$ & (e)$\beta/\mu=2.0$&(f)$\beta/\mu=4.0$ \\ 
     \end{tabular}
    
    \caption{Density plots of the spectral function $A(k_x, k_y)$. Top row illustrates the effect of doping ($\chi = 0.4, 1.0, 2.3$) at $\beta/\mu = 0$, while the bottom row shows the effect of momentum relaxation ($\beta/\mu = 0.4, 2.0, 4.0$) at $\chi = 1.0$. All calculations use $m_\psi = 0$, $q = 1$, and $T/\mu = 0.009$.}
    \label{chibeta-effect}
\end{figure}

We begin by establishing the reference fermionic spectrum in the absence of additional non-minimal interaction channels. 
Figure~\ref{chibeta-effect} shows the zero-frequency spectral density $A(k_x,k_y)$ under variations of the doping parameter $\chi$ and the momentum-relaxation strength $\beta/\mu$. 
The upper row, Figures~\ref{chibeta-effect}(\subref{fig:chi0.4})--\ref{chibeta-effect}(\subref{fig:chi2.3}), displays the effect of doping at fixed $\beta/\mu=0$. 
For small and intermediate values of $\chi$, the spectral weight is mainly concentrated around a closed Fermi ring, while increasing $\chi$ broadens the distribution and weakens the ring structure. 
The lower row, Figures~\ref{chibeta-effect}(\subref{fig:beta0})--\ref{chibeta-effect}(\subref{fig:beta8}), shows the effect of momentum relaxation at fixed $\chi=1.0$. 
Increasing $\beta/\mu$ similarly broadens the spectral weight and reduces the sharpness of the low-energy excitation. 
These reference spectra provide the baseline for identifying the spectral reconstruction induced by the interaction channels discussed below.

The background parameters are chosen according to the structure of each coupling.
For the dipole coupling $p_1$, we work in the clean limit $\beta/\mu=0$
to isolate the dipole-induced spectral response from the effects of momentum relaxation.
By contrast, the $p_2$ and $p_3$ channels explicitly depend on the axion background
and vanish when $\beta=0$.
We therefore fix $\beta/\mu=0.4$ when studying these two channels,
ensuring that the corresponding interaction terms remain nonzero
while retaining finite momentum relaxation.

\subsection{Interaction-Induced Spectral Reconstruction}

\begin{figure}[htb!]
\captionsetup[subfigure]{labelformat=parens}
    \centering
    
    
    \begin{subfigure}[b]{0.16\textwidth}
        \centering
        \includegraphics[width=\textwidth]{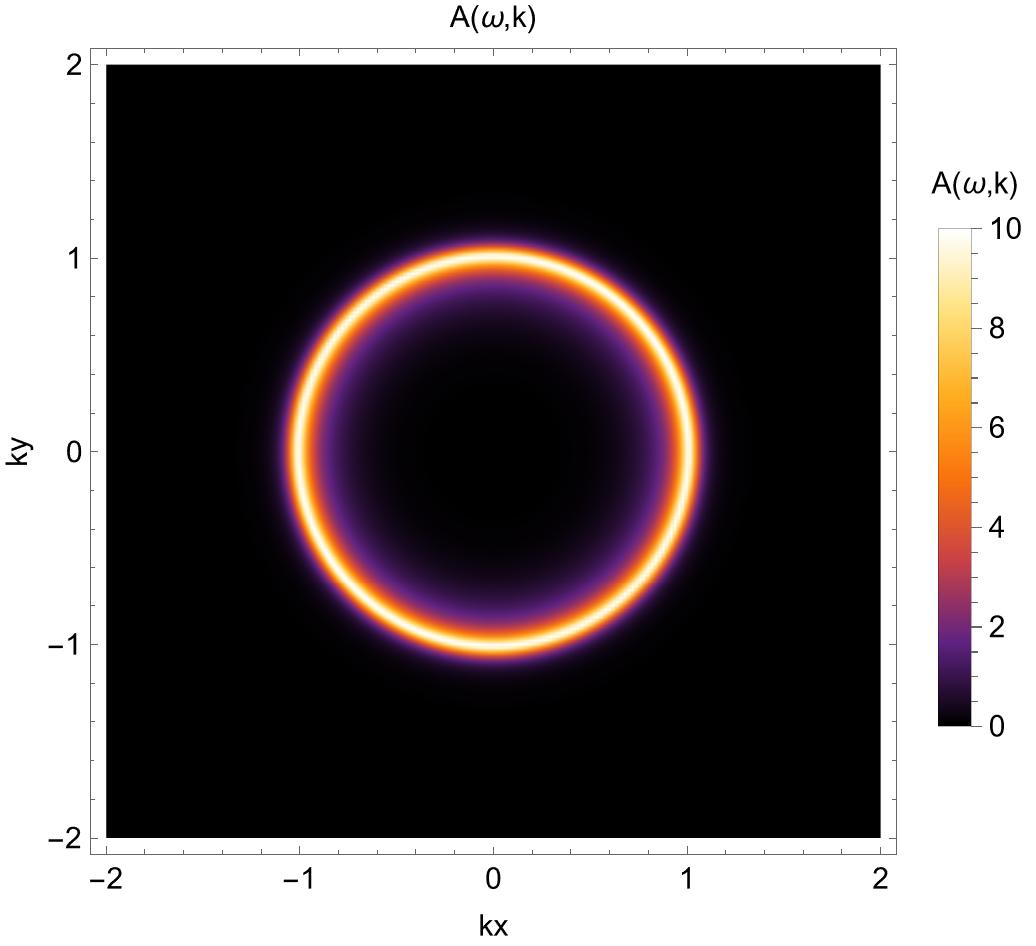}
        \caption{$p_1=2$}
        \label{fig:p1_2}       
    \end{subfigure}\hfill
    \begin{subfigure}[b]{0.16\textwidth}
        \centering
        \includegraphics[width=\textwidth]{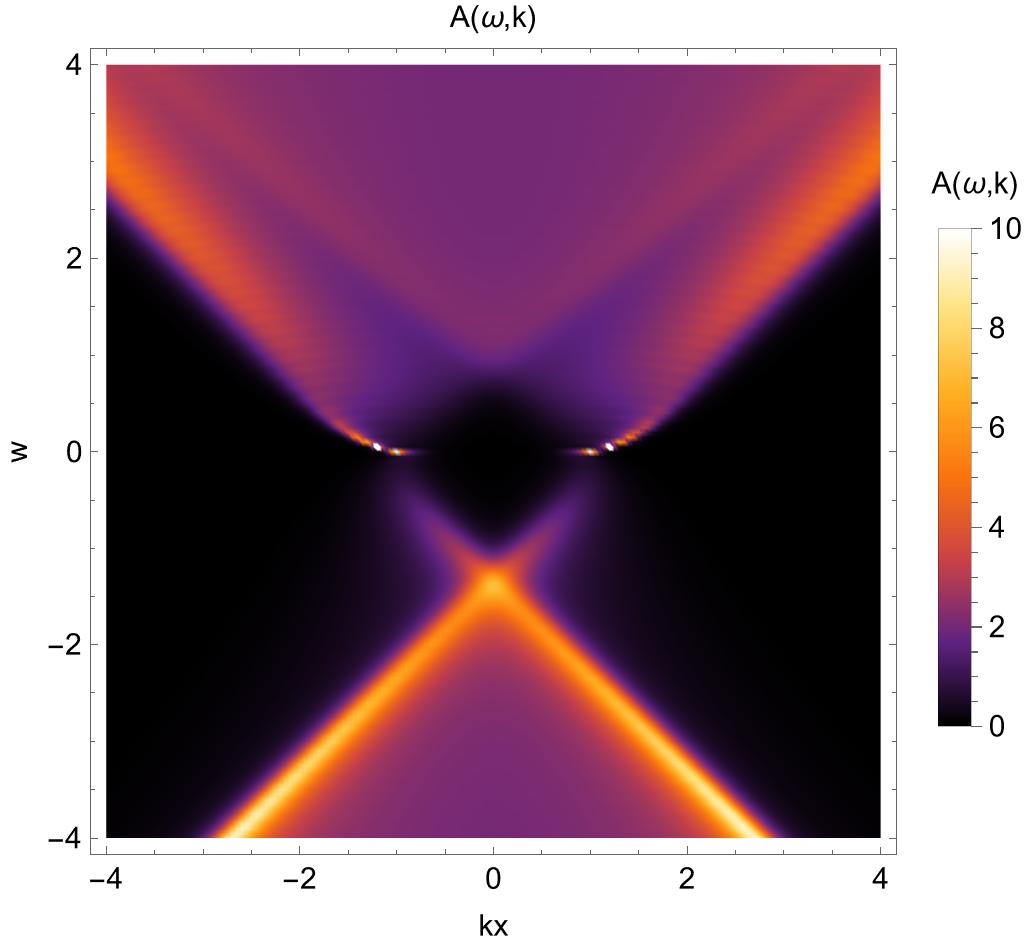}
        \caption{$p_1=2$}
        \label{fig:p1_2_band}  
    \end{subfigure}\hfill
    \begin{subfigure}[b]{0.16\textwidth}
        \centering
        \includegraphics[width=\textwidth]{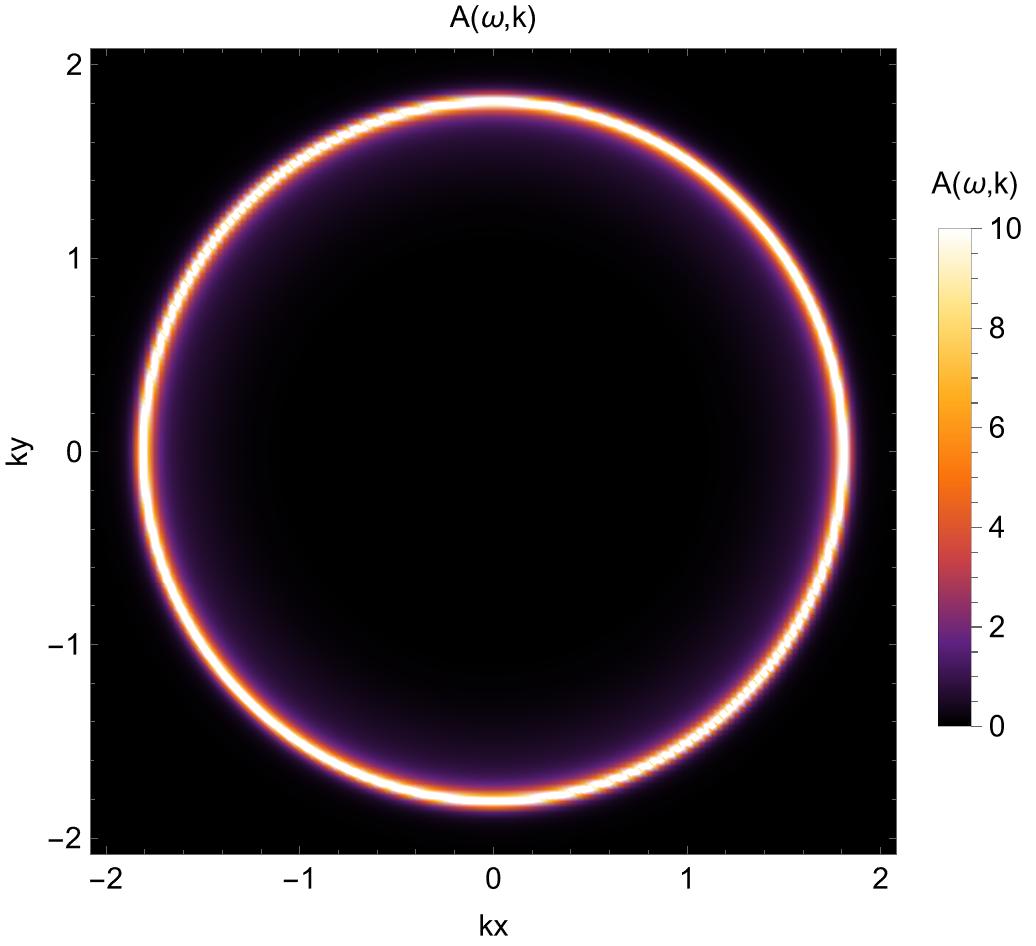}
        \caption{$p_1=4$}
        \label{fig:p1_4}       
    \end{subfigure}\hfill
    \begin{subfigure}[b]{0.16\textwidth}
        \centering
        \includegraphics[width=\textwidth]{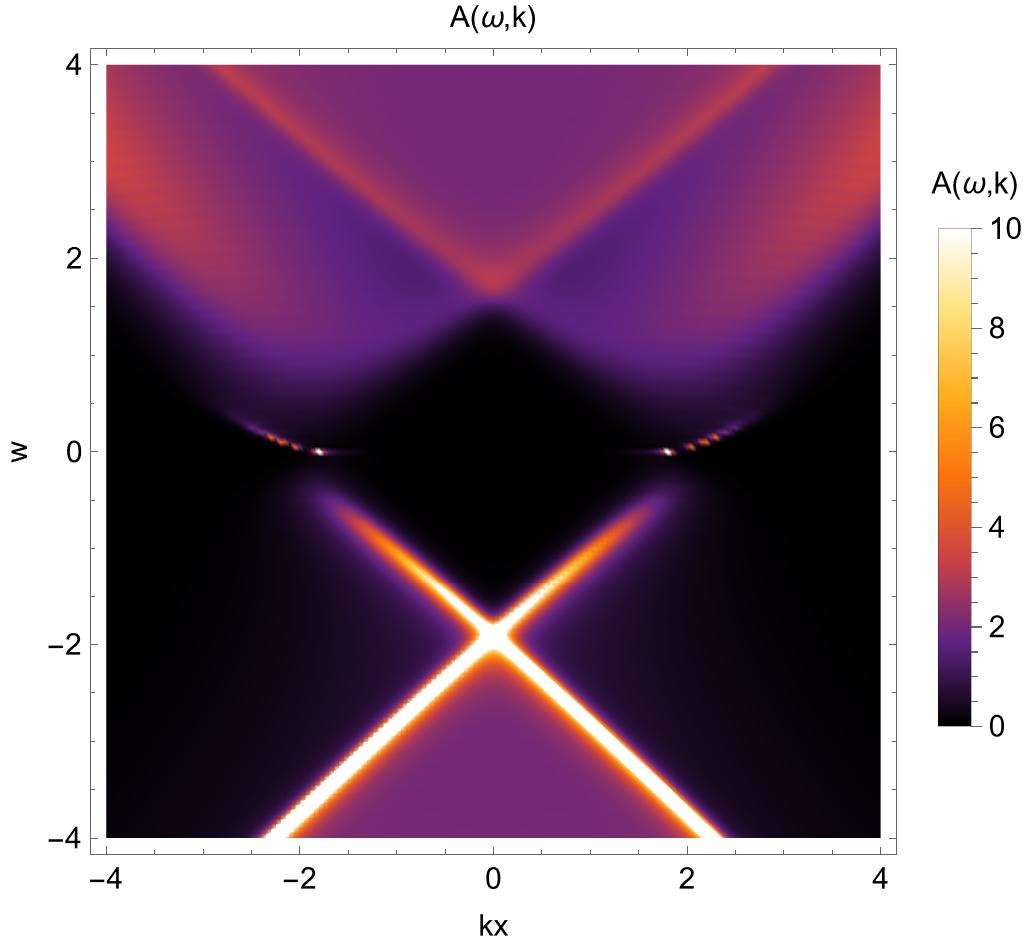}
        \caption{$p_1=4$}
        \label{fig:p1_4_band}  
    \end{subfigure}\hfill
    \begin{subfigure}[b]{0.16\textwidth}
        \centering
        \includegraphics[width=\textwidth]{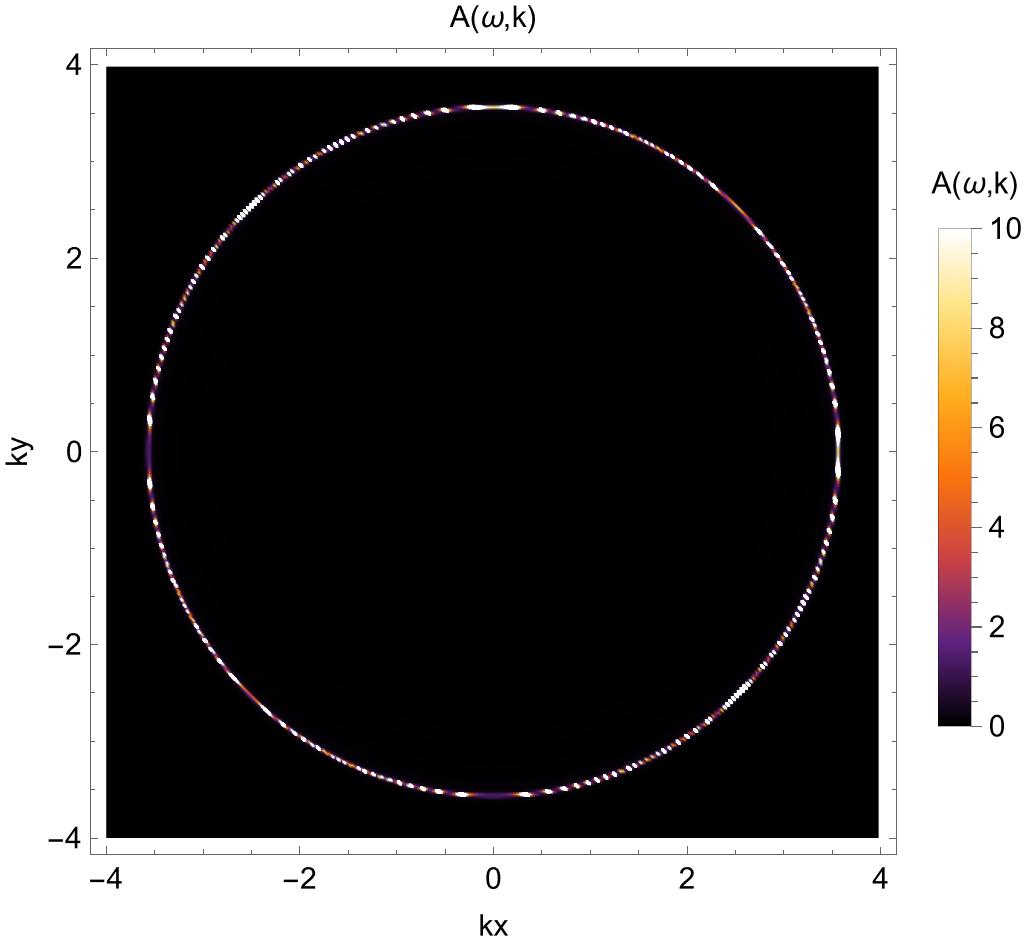}
        \caption{$p_1=8$}
        \label{fig:p1_8}       
    \end{subfigure}\hfill
    \begin{subfigure}[b]{0.16\textwidth}
        \centering
        \includegraphics[width=\textwidth]{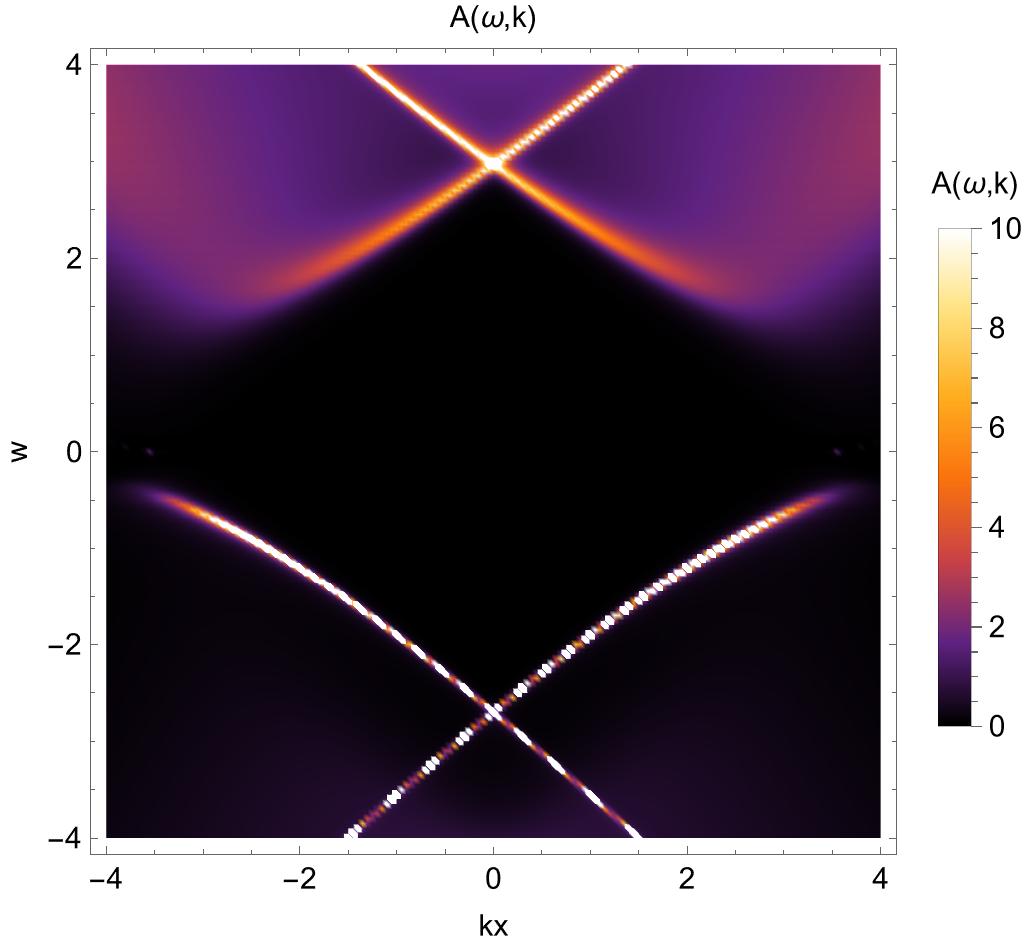}
        \caption{$p_1=8$}
        \label{fig:p1_8_band}  
    \end{subfigure}

    \caption{Density plots of the spectral function $A(k_x, k_y)$ at $\omega=0$ (odd columns) and dispersion relations $A(\omega, k_x)$ along $k_y=0$ (even columns). Fixed parameters are $m_\psi = 0$, $q = 1$, $\chi = 1.0$, and $T/\mu = 0.009$ in the clean limit ($\beta/\mu = 0$). Panels (a)--(f) present the spectral evolution with dipole coupling strengths $p_1 = 2, 4, \text{and } 8$. }
    \label{p1-eta-effect}
\end{figure}

We first examine the dipole coupling $p_1$, which serves as a
comparison channel for the axion-related interactions. 
Figure~\ref{p1-eta-effect} shows the zero-frequency spectral density $A(k_x,k_y)$ and the corresponding dispersion relation $A(\omega,k_x)$ for $p_1=2,4,8$. 
As $p_1$ increases, the zero-energy spectral weight remains distributed along a closed Fermi ring, while the ring radius increases. 
Thus, the dipole coupling expands the Fermi surface but does not break it into open segments. 
The dispersion plots show an asymmetric redistribution of spectral weight around the Fermi level, indicating a soft reconstruction of the low-energy spectrum.

\begin{figure}[htbp]
    \centering
    \begin{subfigure}[b]{0.50\textwidth}
        \centering
        \includegraphics[width=\textwidth]{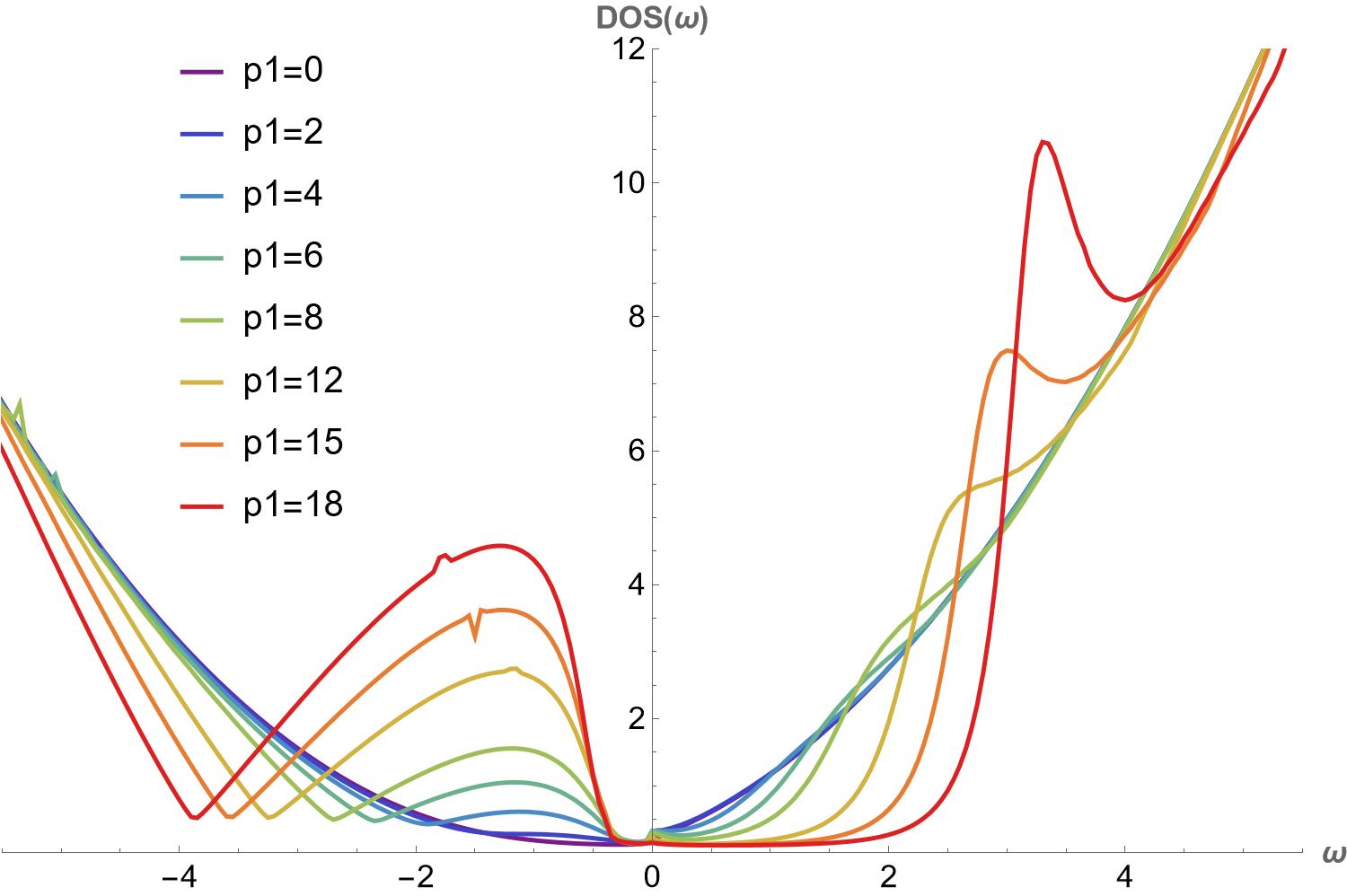}
        \caption{}
        \label{fig:p1_DOS_omega}
    \end{subfigure}

    \vspace{-0.10cm}

\begin{subfigure}[t]{0.43\textwidth}
    \centering
    \includegraphics[height=4.0cm, keepaspectratio]{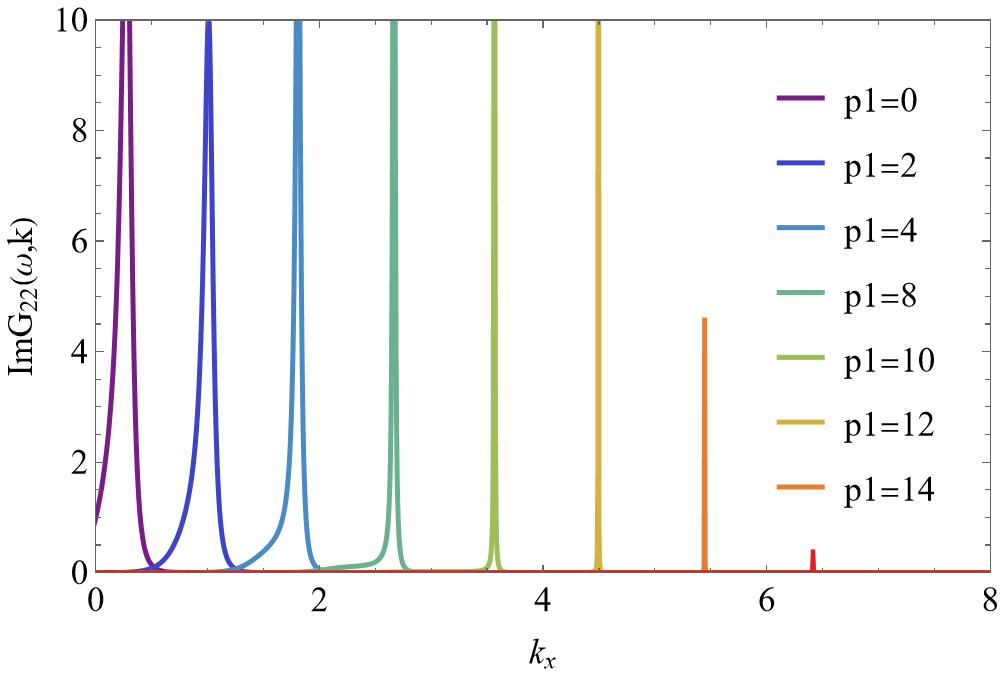}
    \caption{}
    \label{fig:p1_kx_cuts}
\end{subfigure}
\hspace{0.03\textwidth}
\begin{subfigure}[t]{0.43\textwidth}
    \centering
    \includegraphics[height=4.2cm, keepaspectratio]{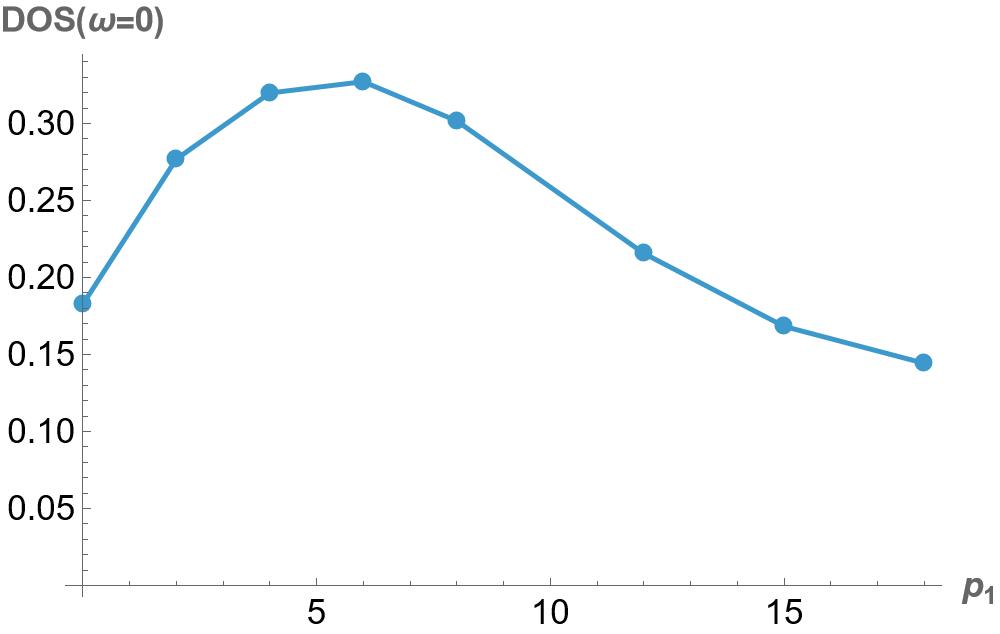}
    \caption{}
    \label{fig:p1_DOS_zero}
\end{subfigure}

    \caption{
    Spectral functions for different values of the  dipole coupling $p_1$.
(a) Momentum-integrated density of states $\mathrm{DOS}(\omega)$.
(b) Zero-energy momentum cuts $ImG_{22}(k_x)$ at $\omega=0$.
(c) Zero-frequency density of states $\mathrm{DOS}(0)$ as a function of $p_1$.
Other parameters are fixed as $m_\psi=0$, $q=1$, $\chi=1.0$, $T/\mu=0.009$, and $\beta/\mu=0$.
The finite and non-monotonic $\mathrm{DOS}(0)$ suggests a soft low-energy suppression rather than a hard gap.
}
    \label{fig:p1_diagnostics}
\end{figure}

Exploiting the rotational symmetry of the clean background, we
evaluate the density of states as
\begin{align}
\mathrm{DOS}(\omega)
= \frac{1}{2\pi}\int_0^{k_{\max}} dk\,k\,A(\omega,k),
\end{align}
where $k_{\max}$ is the numerical momentum cutoff.

This behavior is further clarified by the density of states and the momentum cuts shown in Figure~\ref{fig:p1_diagnostics}. 
The momentum-integrated density of states $\mathrm{DOS}(\omega)$ in Figure~\ref{fig:p1_diagnostics}(\subref{fig:p1_DOS_omega}) develops a pronounced particle-hole asymmetry as $p_1$ increases. 
Meanwhile, the zero-energy momentum cuts \(\mathrm{Im}G_{22}(k_x)\) at \(\omega=0\) in Figure~\ref{fig:p1_diagnostics}(\subref{fig:p1_kx_cuts}) show that the Fermi momentum is pushed outward with increasing $p_1$. 
The zero-frequency density of states $\mathrm{DOS}(0)$ in Figure~\ref{fig:p1_diagnostics}(\subref{fig:p1_DOS_zero}) exhibits a non-monotonic dependence on $p_1$: it first increases at intermediate coupling and then decreases in the stronger-coupling regime. 
This non-monotonic behavior can be understood as a competition between
the outward shift of the Fermi surface and the suppression of low-energy
spectral weight. Since the momentum-integrated DOS carries the
two-dimensional phase-space factor $k$, the outward shift of the Fermi
surface can initially enhance $\mathrm{DOS}(0)$, whereas at stronger
coupling the reduction of low-energy spectral weight becomes dominant.

Although $\mathrm{DOS}(0)$ remains finite over the parameter range
considered, the spectrum develops a soft and particle-hole-asymmetric
reconstruction reminiscent of the Mott-like behavior discussed in
Ref.~\cite{Ghorai:2024nxs}. Within the present parameter range and
numerical resolution, however, no clean hard Mott gap is resolved.

In the bulk Dirac equation, the Pauli-type interaction modifies the
radial dynamics of the probe fermion and thereby changes the momentum
location and spectral weight of the low-energy excitations. This is
consistent with the known role of dipole couplings in holographic
fermion systems, where they can produce substantial spectral-weight
redistribution and low-energy suppression~\cite{Ghorai:2024nxs}.
Having established the dipole channel as a soft-reconstruction baseline,
we now turn to the axion-related $p_2$ and $p_3$ interactions, for which
current mixing and the momentum-relaxation sector enter explicitly.

\begin{figure}[htbp]
    \centering
    \captionsetup[subfigure]{labelformat=parens, skip=2pt} %
    \newcommand{\rowheight}{2.5cm}

    \begin{subfigure}[b]{0.16\linewidth}
        \centering
        \includegraphics[width=\linewidth, height=\rowheight, keepaspectratio]{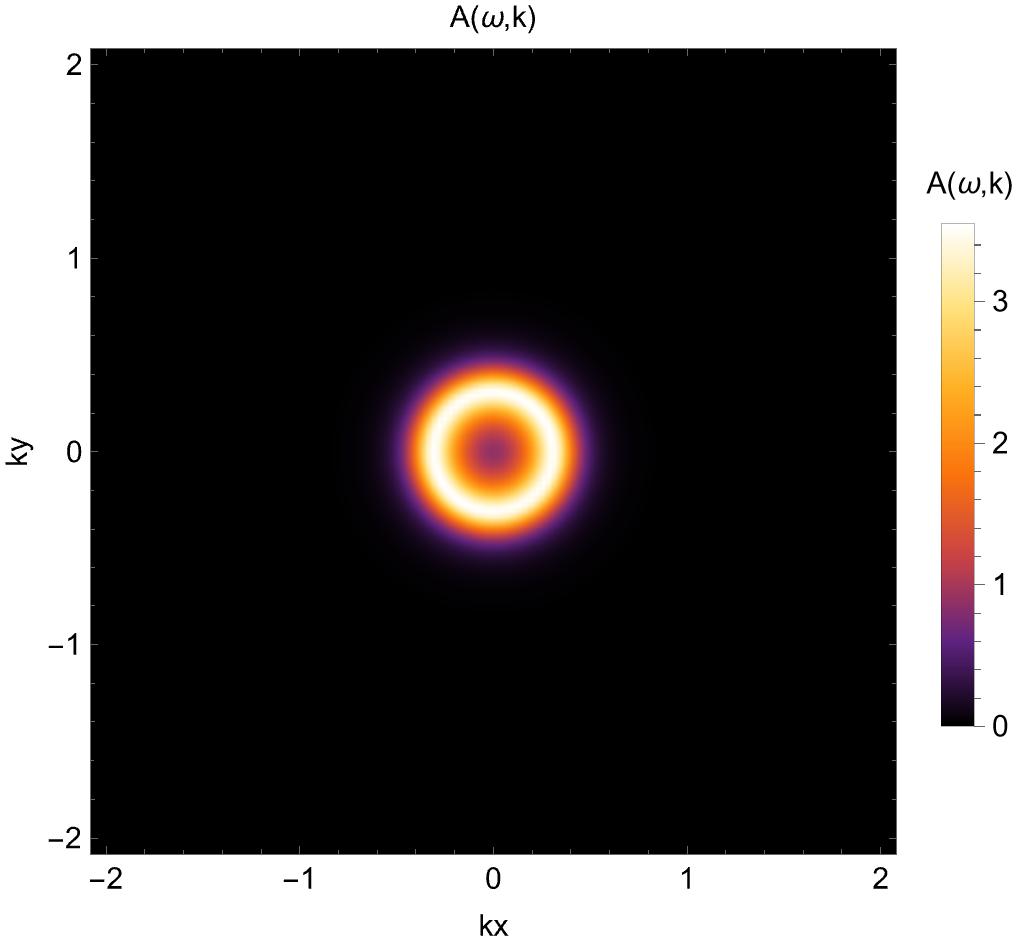}
        \caption{\centering $p_2=0.2$}
        \label{fig:row1_1}
    \end{subfigure}\hfill
    \begin{subfigure}[b]{0.16\linewidth}
        \centering
        \includegraphics[width=\linewidth, height=\rowheight, keepaspectratio]{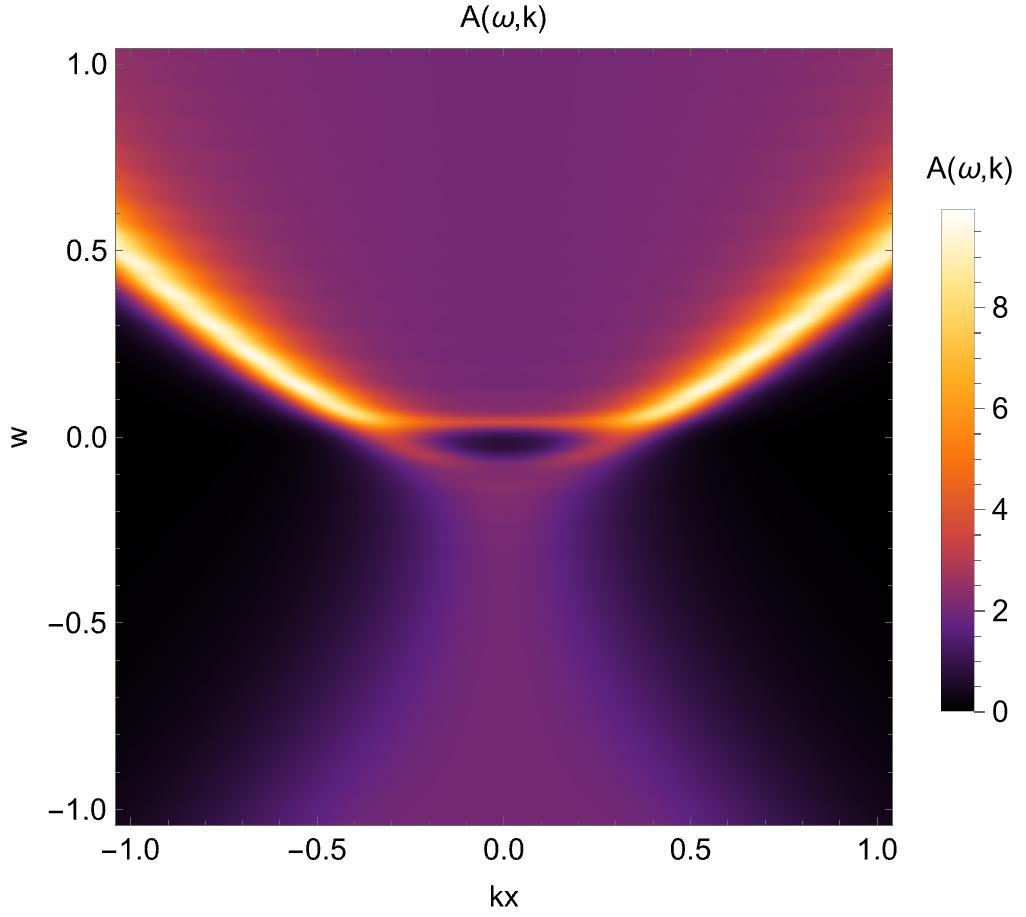}
        \caption{\centering $p_2=0.2$}
        \label{fig:row1_2}
    \end{subfigure}\hfill
    \begin{subfigure}[b]{0.16\linewidth}
        \centering
        \includegraphics[width=\linewidth, height=\rowheight, keepaspectratio]{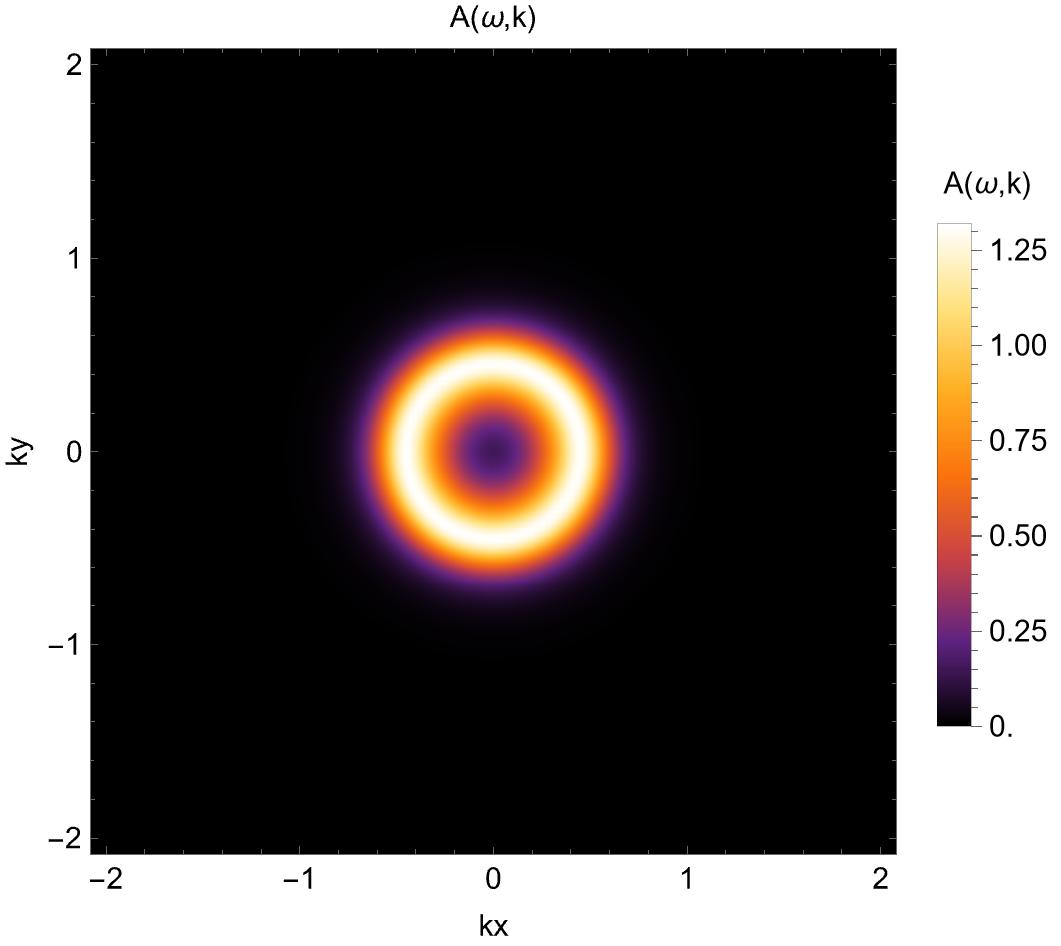}
        \caption{\centering $p_2=0.4$}
        \label{fig:row1_3}
    \end{subfigure}\hfill
    \begin{subfigure}[b]{0.16\linewidth}
        \centering
        \includegraphics[width=\linewidth, height=\rowheight, keepaspectratio]{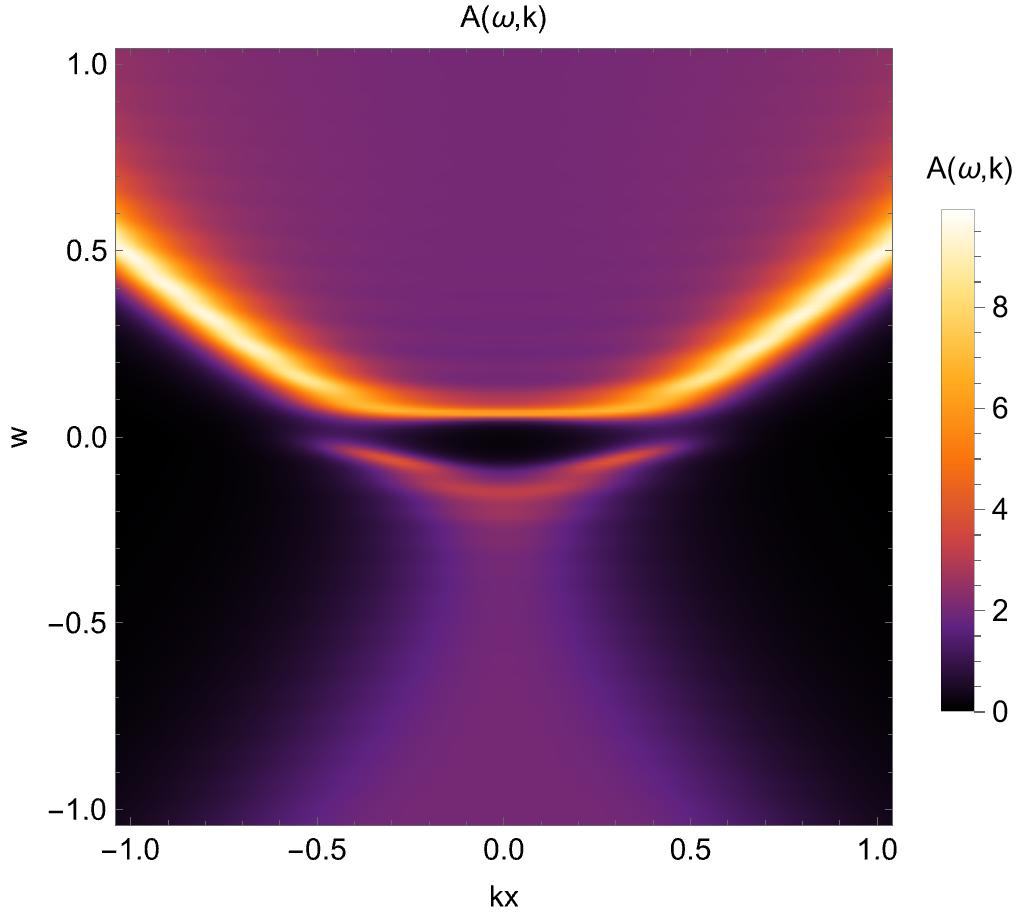}
        \caption{\centering $p_2=0.4$}
        \label{fig:row1_4}
    \end{subfigure}\hfill
    \begin{subfigure}[b]{0.16\linewidth}
        \centering
        \includegraphics[width=\linewidth, height=\rowheight, keepaspectratio]{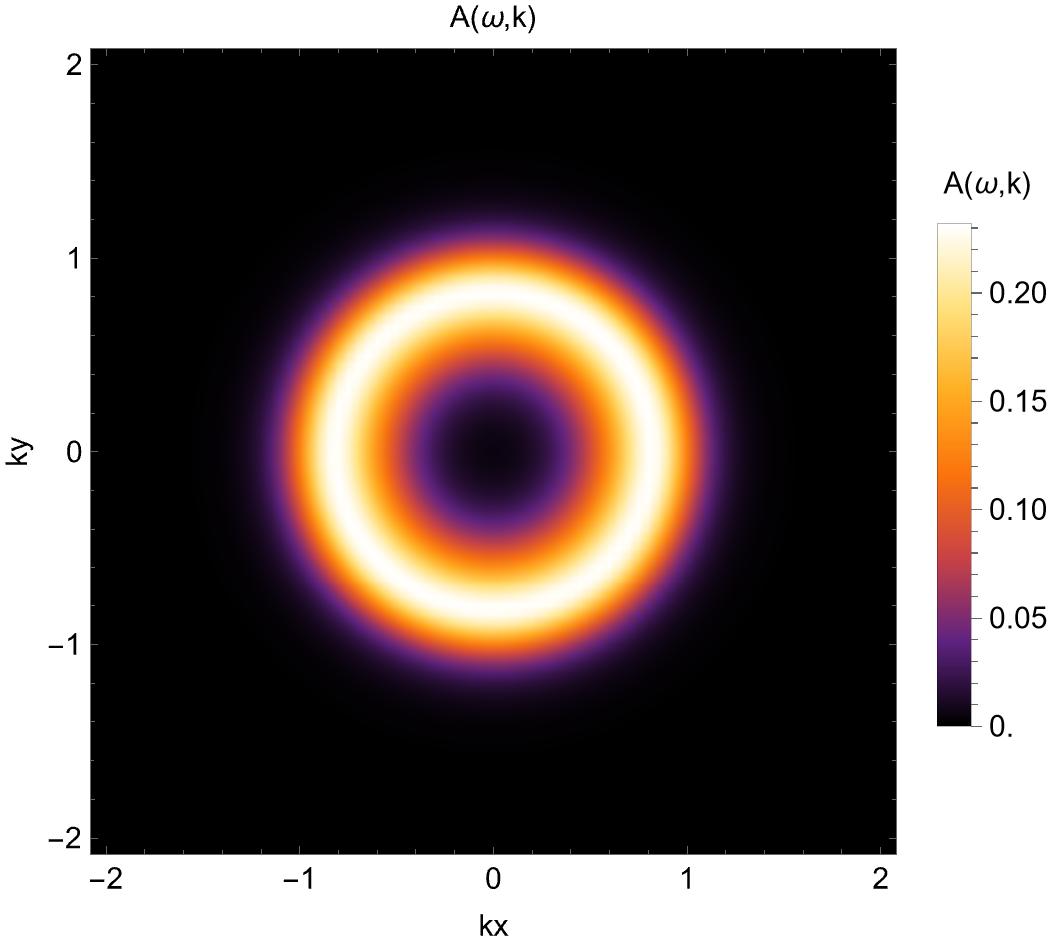}
        \caption{\centering $p_2=0.8$}
        \label{fig:row1_5}
    \end{subfigure}\hfill
    \begin{subfigure}[b]{0.16\linewidth}
        \centering
        \includegraphics[width=\linewidth, height=\rowheight, keepaspectratio]{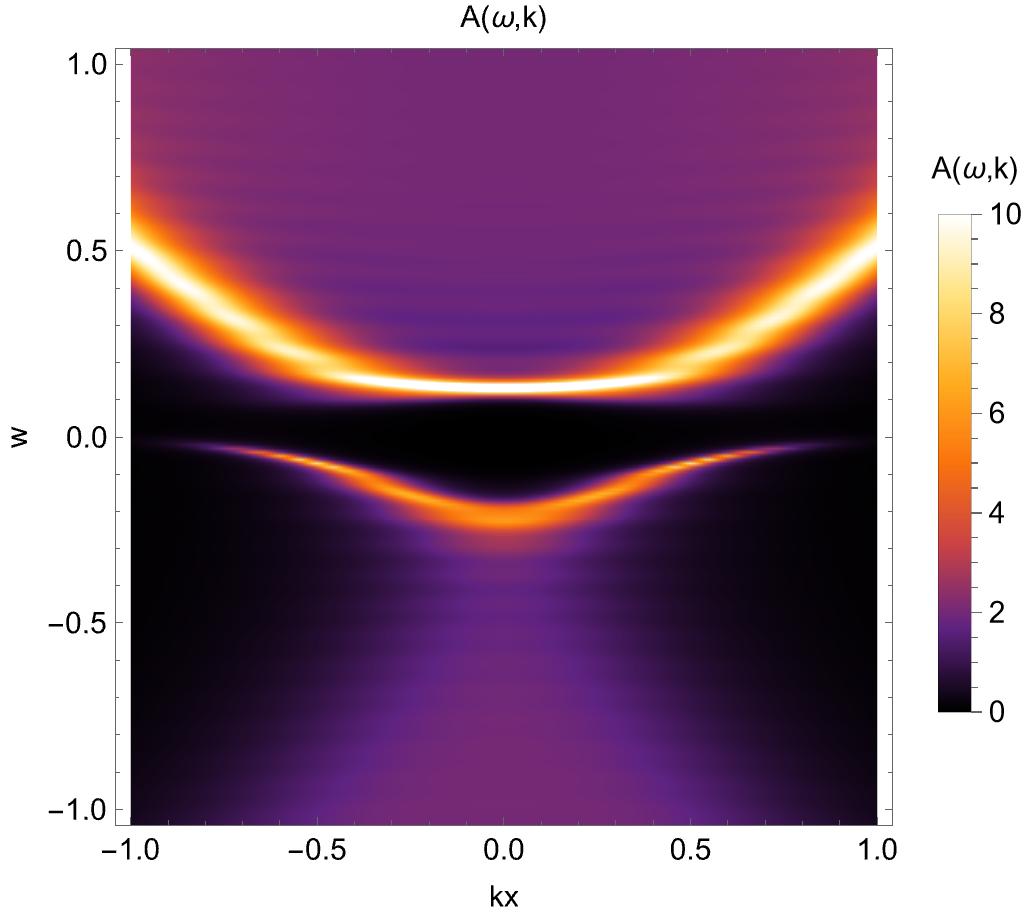}
        \caption{\centering $p_2=0.8$}
        \label{fig:row1_6}
    \end{subfigure}

    \vspace{0.8em} 

    
    \begin{subfigure}[b]{0.16\linewidth}
        \centering
        \includegraphics[width=\linewidth, height=\rowheight, keepaspectratio]{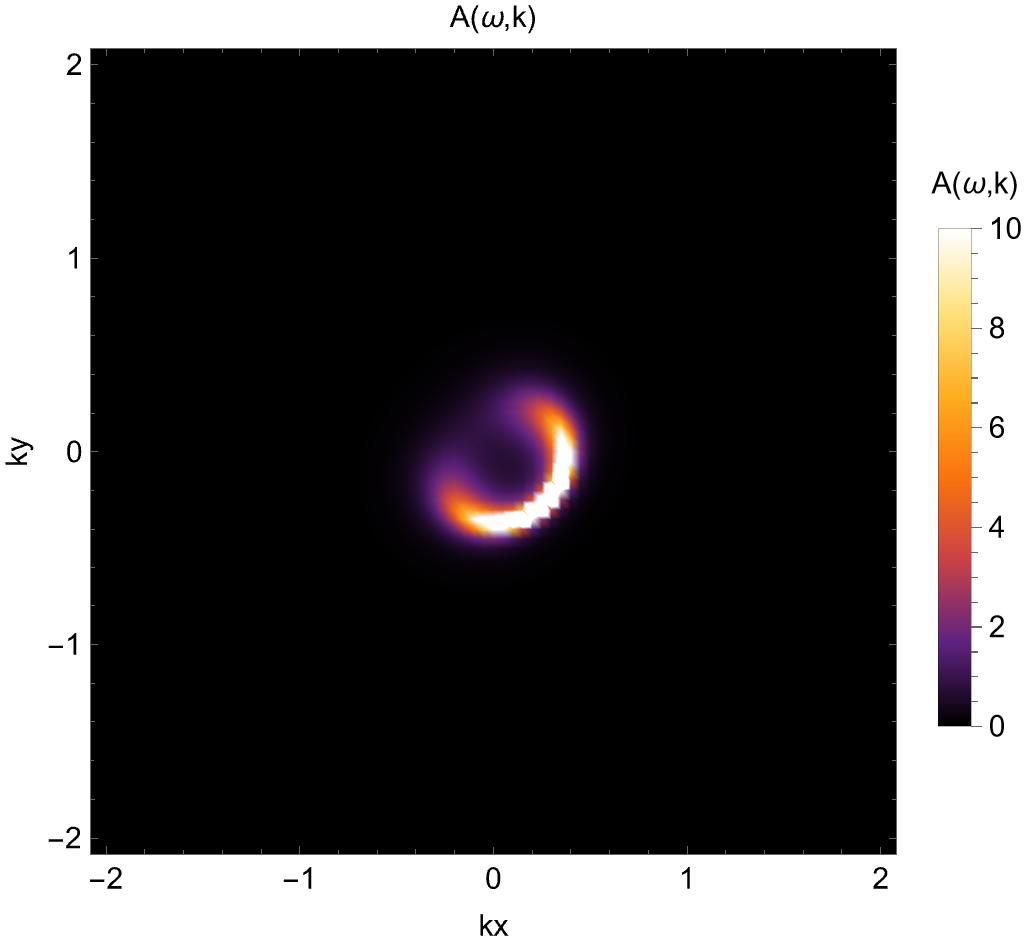}
        \caption{\centering $p_3=0.2$}
        \label{fig:row2_1}
    \end{subfigure}\hfill
    \begin{subfigure}[b]{0.16\linewidth}
        \centering
        \includegraphics[width=\linewidth, height=\rowheight, keepaspectratio]{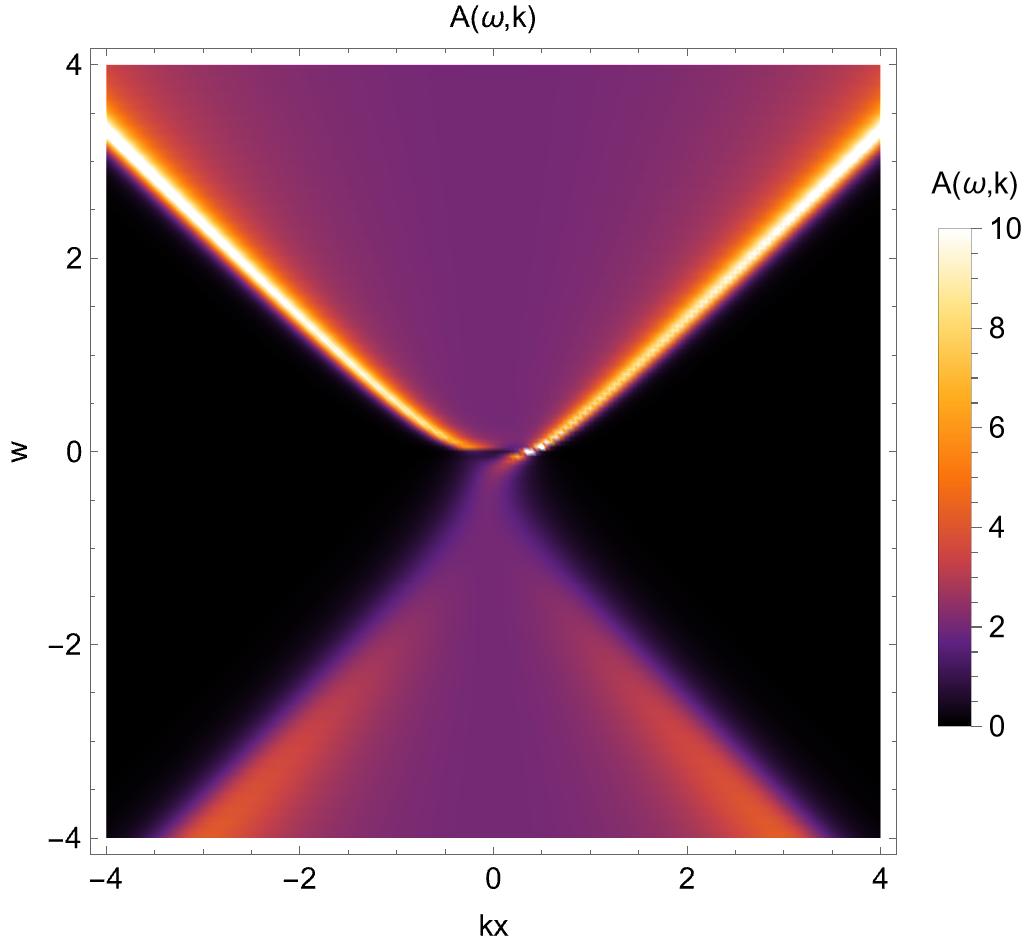}
        \caption{\centering $p_3=0.2$}
        \label{fig:row2_2}
    \end{subfigure}\hfill
    \begin{subfigure}[b]{0.16\linewidth}
        \centering
        \includegraphics[width=\linewidth, height=\rowheight, keepaspectratio]{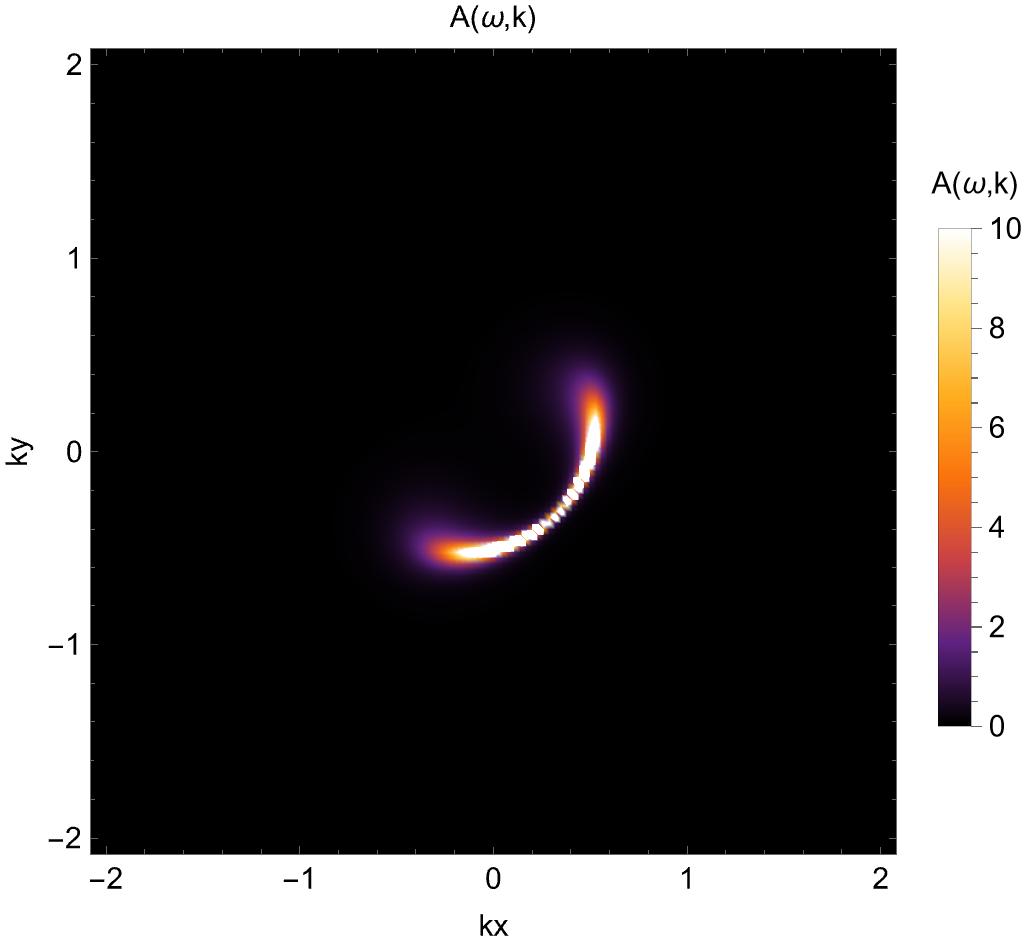}
        \caption{\centering $p_3=0.5$}
        \label{fig:row2_3}
    \end{subfigure}\hfill
    \begin{subfigure}[b]{0.16\linewidth}
        \centering
        \includegraphics[width=\linewidth, height=\rowheight, keepaspectratio]{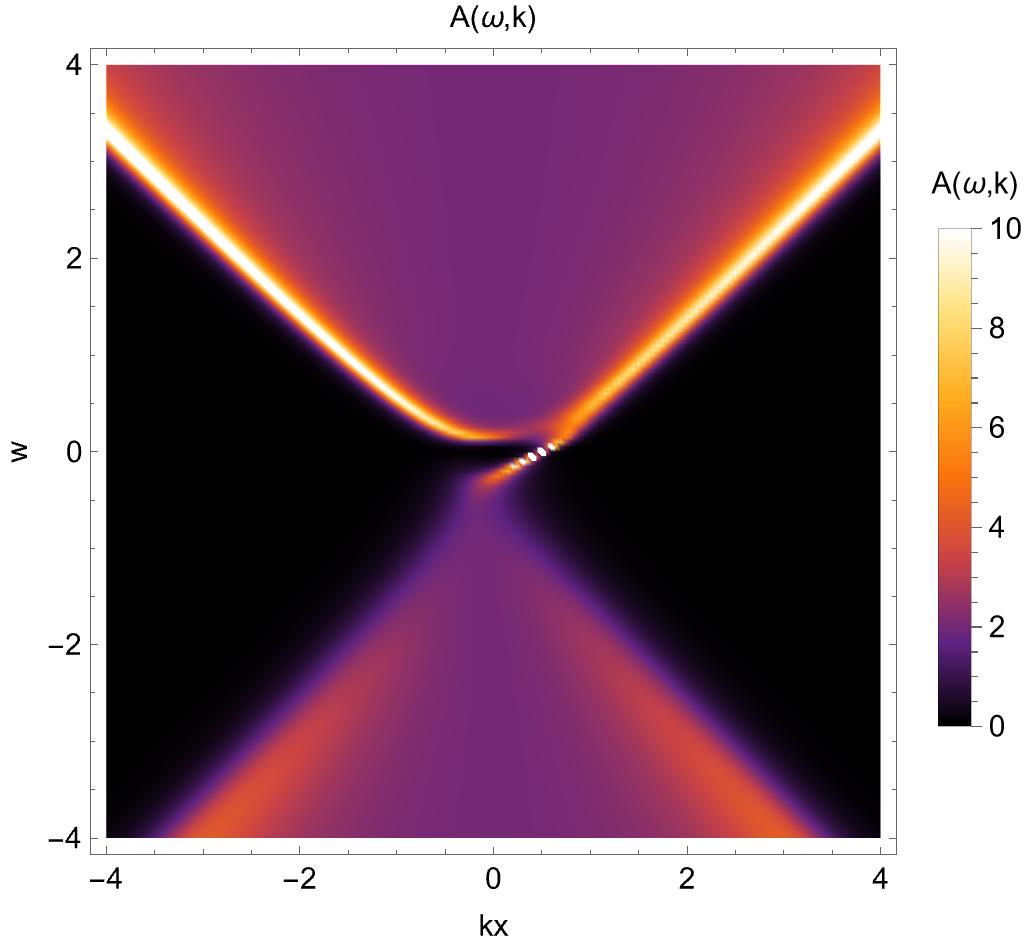}
        \caption{\centering $p_3=0.5$}
        \label{fig:row2_4}
    \end{subfigure}\hfill
    \begin{subfigure}[b]{0.16\linewidth}
        \centering
        \includegraphics[width=\linewidth, height=\rowheight, keepaspectratio]{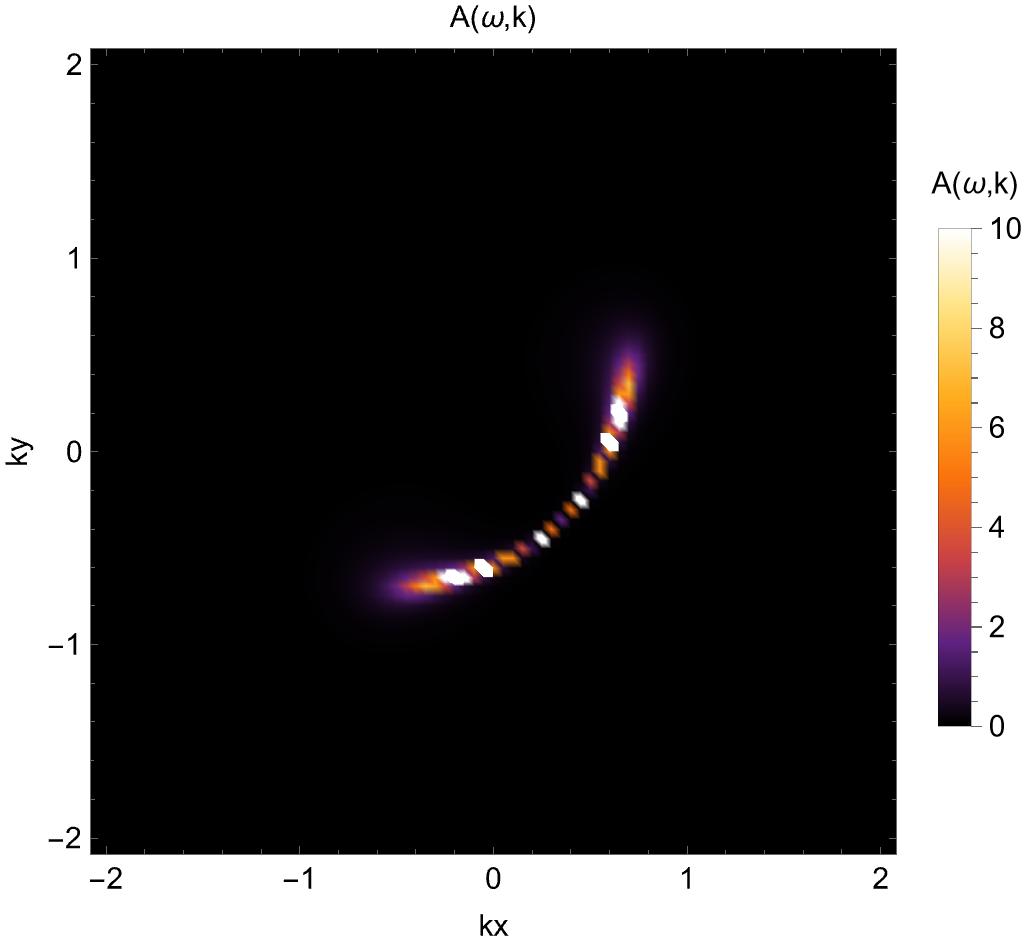}
        \caption{\centering $p_3=0.8$}
        \label{fig:row2_5}
    \end{subfigure}\hfill
    \begin{subfigure}[b]{0.16\linewidth}
        \centering
        \includegraphics[width=\linewidth, height=\rowheight, keepaspectratio]{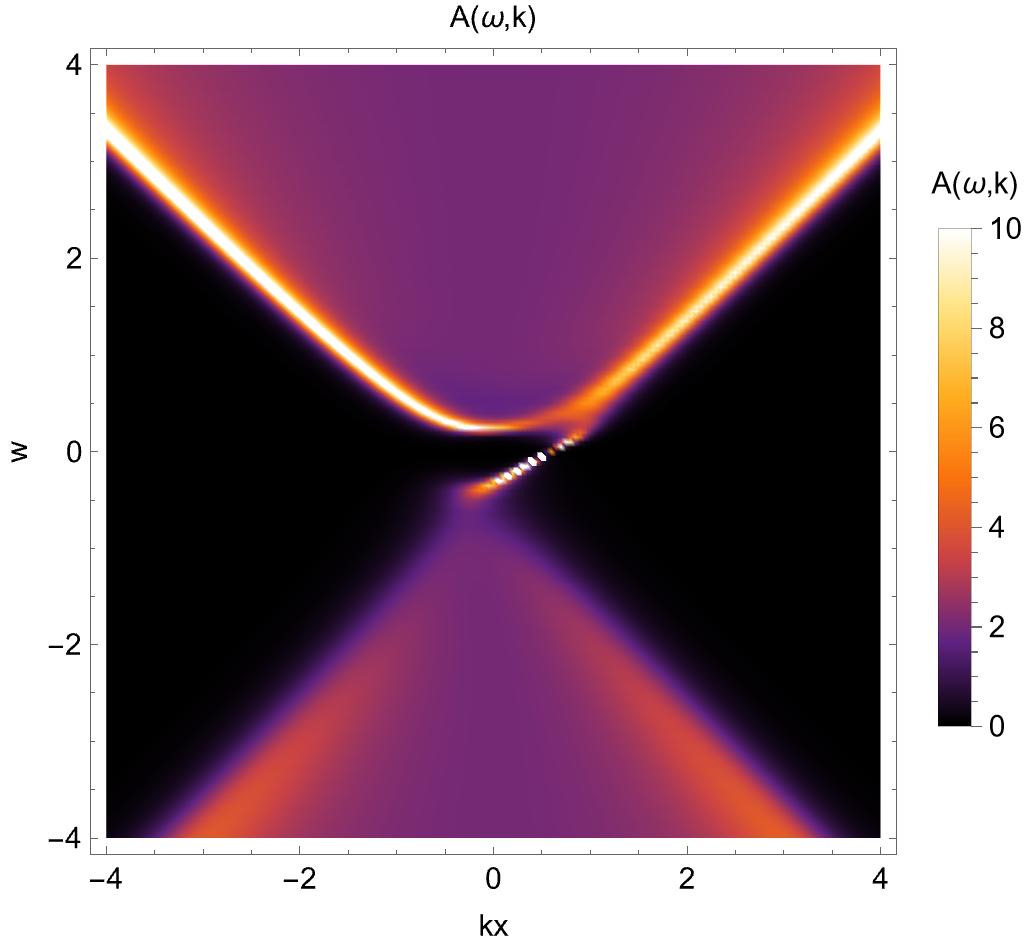}
        \caption{\centering $p_3=0.8$}
        \label{fig:row2_6}
    \end{subfigure}

    \caption{Density plots of the spectral function $A(k_x, k_y)$ at $\omega=0$ (odd columns) and dispersion relations $A(\omega, k_x)$ along $k_y=0$ (even columns). Fixed parameters are $m_\psi = 0$, $q = 1$, $\chi = 1.0$, and $T/\mu = 0.009$ with finite momentum relaxation ($\beta/\mu = 0.4$). The top row (a)--(f) presents the spectral evolution driven by the tensor coupling $p_2$ set to $0.2, 0.4$, and $0.8$, showing the band flattening effect. The bottom row (g)--(l) corresponds to the vector coupling $p_3$ set to $0.2, 0.5$, and $0.8$, respectively, showing arc-like spectral redistribution.}
    \label{fig:full_comparison}
\end{figure}

\begin{figure}[htbp]
    \centering
    \captionsetup[subfigure]{labelformat=parens}

    \newcommand{\gridfigheight}{4.5cm} 

 \begin{subfigure}[b]{0.48\linewidth}  
        \centering
        \includegraphics[width=\linewidth, height=\gridfigheight, keepaspectratio]{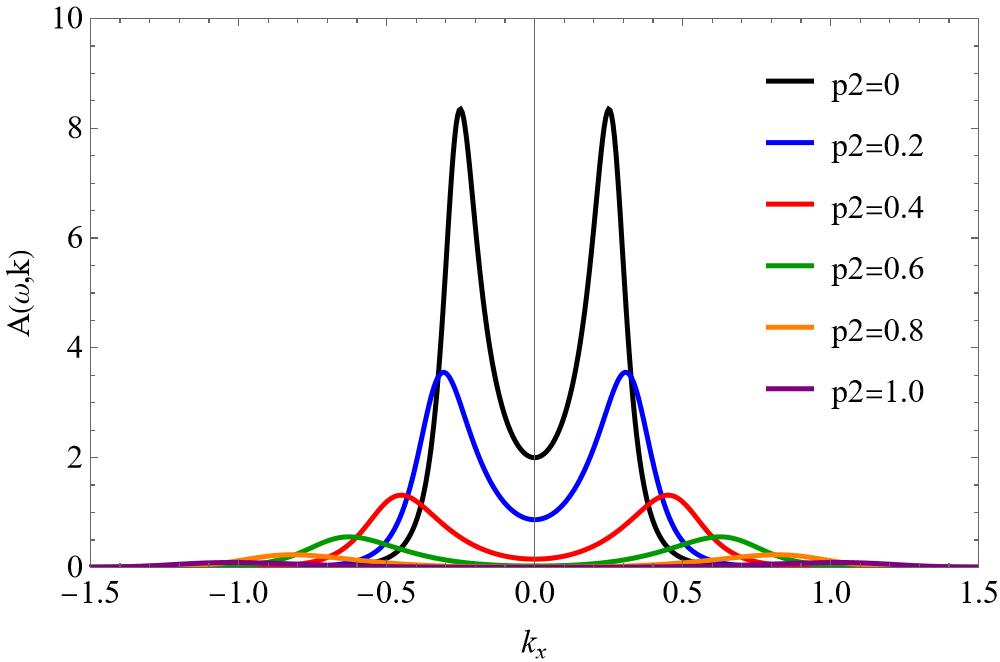}
        \caption{ }
        \label{fig:kxA1-p2}
    \end{subfigure}\hfill 
    \begin{subfigure}[b]{0.48\linewidth}  
        \centering
        \includegraphics[width=\linewidth, height=\gridfigheight, keepaspectratio]{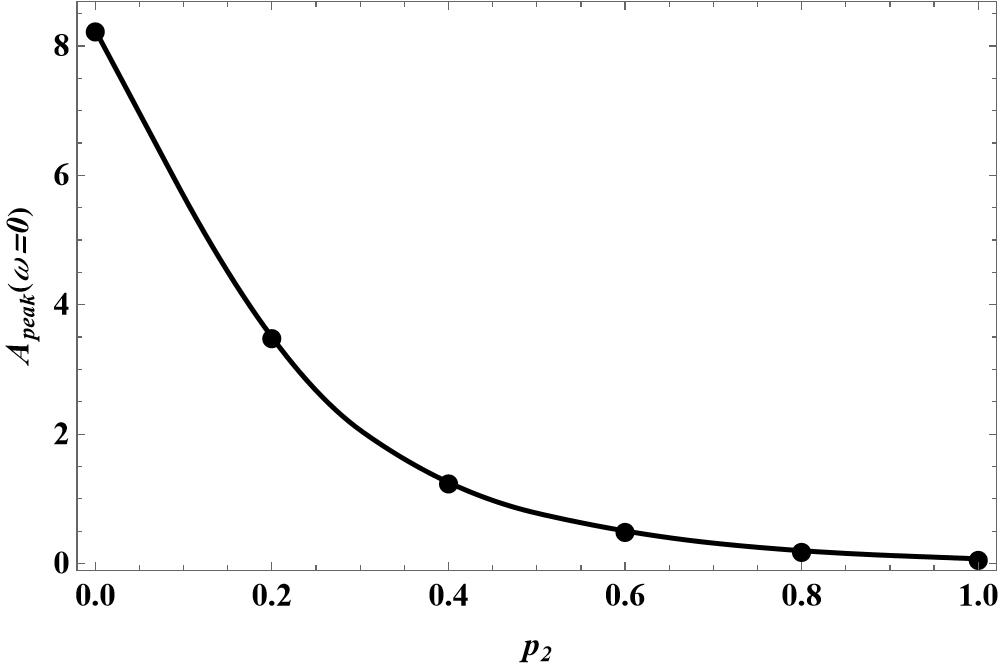}
        \caption{}
        \label{fig:A_p2} 
    \end{subfigure}
 \vspace{1em} 

    \begin{subfigure}[b]{0.48\linewidth}
        \centering
        \includegraphics[width=\linewidth, height=\gridfigheight, keepaspectratio]{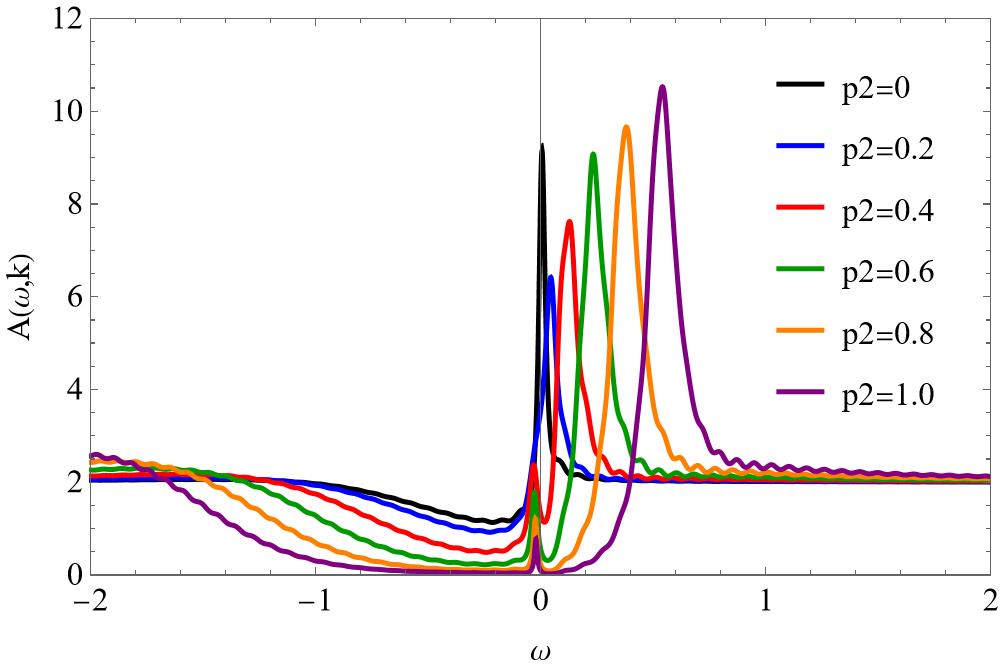}
        \caption{}
        \label{fig:Aw-p2}
    \end{subfigure}\hfill
    \begin{subfigure}[b]{0.48\linewidth}
        \centering
        \includegraphics[width=\linewidth, height=\gridfigheight, keepaspectratio]{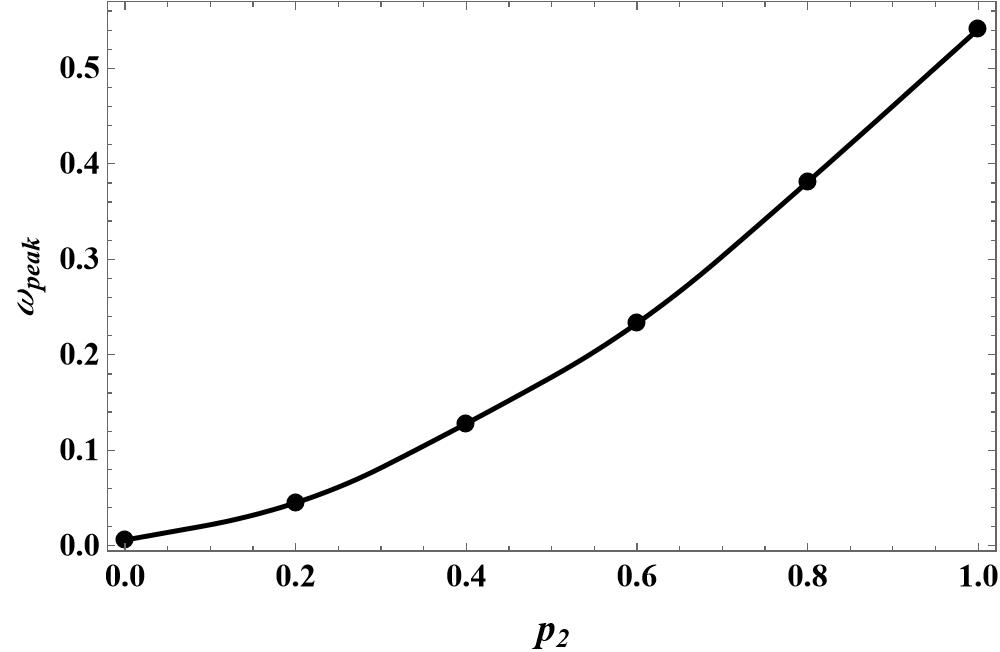}
        \caption{}
        \label{fig:w-p2}
    \end{subfigure}

    \caption{
Spectral function cuts and peak features for the $p_2$ coupling.
Top row: (a) $A(k_x)$ at $\omega=0$ and (b) the zero-energy peak height $A_{\rm peak}$ as a function of $p_2$.
Bottom row: (c) $A(\omega,k_F(p_2))$ evaluated at the corresponding peak momentum and (d) the peak position $\omega_{\rm peak}$ as a function of $p_2$.
The suppression of $A_{\rm peak}$ and the shift of $\omega_{\rm peak}$ indicate that the $p_2$ coupling redistributes spectral weight away from the Fermi level.
The fixed parameters are $m_\psi=0$, $q=1$, $\chi=1.0$, $\beta/\mu=0.4$, and $T/\mu=0.009$.
}
    \label{fig:p2_diagnostics}
\end{figure}

\begin{figure}[htbp]
    \centering
    \captionsetup[subfigure]{labelformat=parens}

    \newcommand{\gridfigheight}{4.5cm} 

    \begin{subfigure}[b]{0.48\linewidth}
        \centering
        \includegraphics[width=\linewidth, height=\gridfigheight, keepaspectratio]{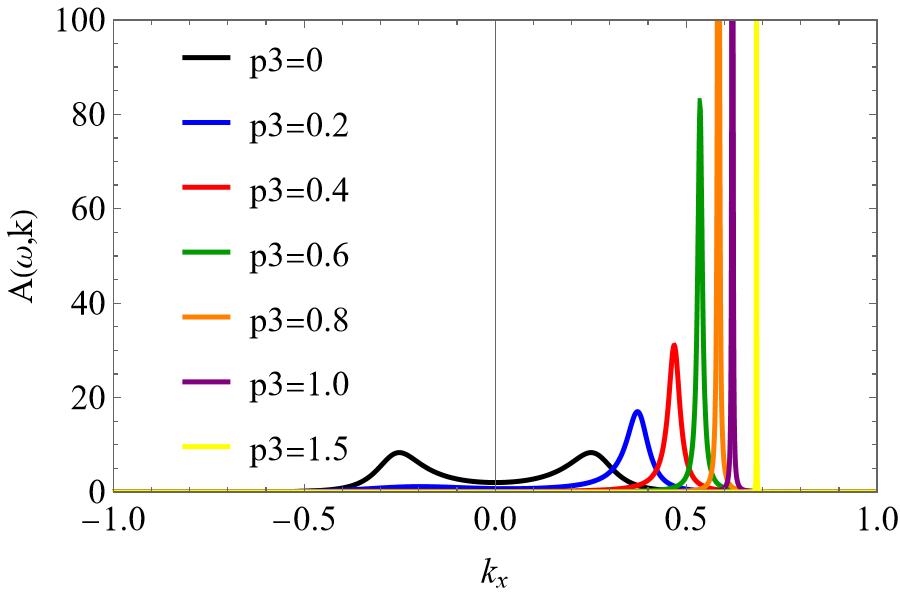}
        \caption{}
      \label{fig:p3_kx_cuts}
    \end{subfigure}\hfill
    \begin{subfigure}[b]{0.48\linewidth}
        \centering
        \includegraphics[width=\linewidth, height=\gridfigheight, keepaspectratio]{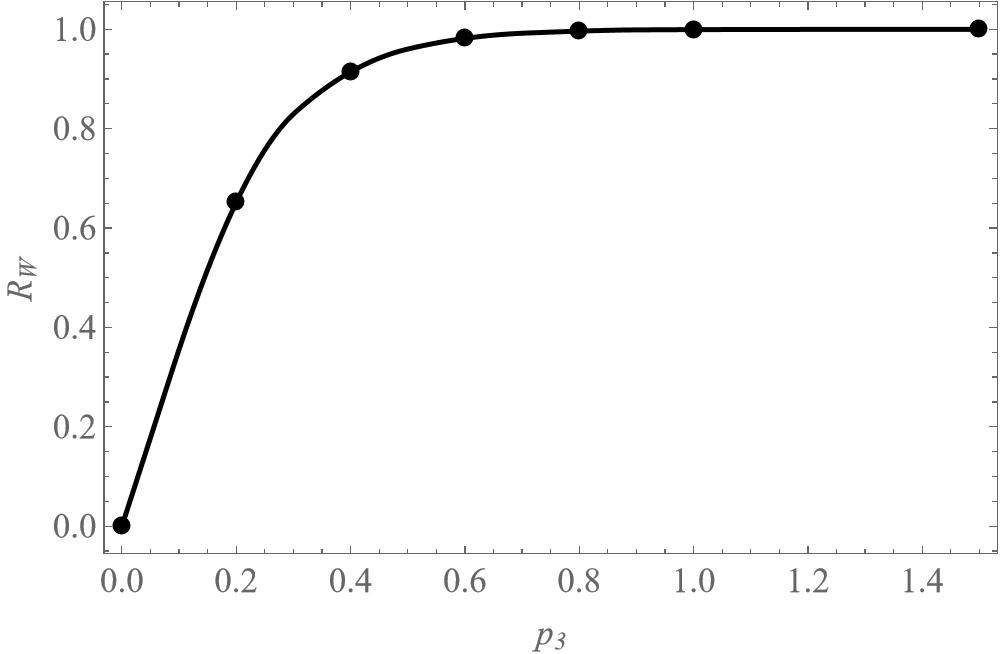}
        \caption{}
        \label{fig:p3_RW}
    \end{subfigure}

    \caption{
Zero-frequency spectral function cuts for the $p_3$ coupling.
(a) $A(k_x)$ at $\omega=0$ for different values of $p_3$.
(b) The integrated asymmetry ratio $R_W$ as a function of $p_3$.
As $p_3$ increases, $R_W$ rapidly approaches unity, showing that the spectral weight becomes concentrated on the positive-momentum branch.
The fixed parameters are $m_\psi=0$, $q=1$, $\chi=1.0$, $\beta/\mu=0.4$, and $T/\mu=0.009$.
}
    \label{p3-eta-akx-w}
\end{figure}

We begin with the tensor coupling $p_2$. 
Unlike the dipole channel, the $p_2$ interaction is activated only in the presence of a finite axion background; therefore, we fix $\beta/\mu=0.4$ in this part. 
The upper row of Figure~\ref{fig:full_comparison} shows the evolution of the zero-frequency spectral density $A(k_x,k_y)$ and the dispersion relation $A(\omega,k_x)$ for $p_2=0.2,0.4,0.8$. 
The density plots in Figures~\ref{fig:full_comparison}(\subref{fig:row1_1}), \ref{fig:full_comparison}(\subref{fig:row1_3}), and \ref{fig:full_comparison}(\subref{fig:row1_5}) show a progressive
reconstruction of the zero-frequency spectral distribution as $p_2$
increases. The suppression of the zero-energy peak is quantified more
directly below in Figure~\ref{fig:p2_diagnostics}.
The corresponding dispersion plots in Figures~\ref{fig:full_comparison}(\subref{fig:row1_2}), \ref{fig:full_comparison}(\subref{fig:row1_4}), and \ref{fig:full_comparison}(\subref{fig:row1_6}) reveal a visible deformation of the low-energy spectral branches. 
In particular, the bottom of the upper branch becomes increasingly flat, while the spectral weight near the original Fermi level is reduced. 
These features indicate that the $p_2$ channel produces low-energy spectral suppression accompanied by a flat-bottom structure in the upper branch.

The corresponding peak features are shown in Figure~\ref{fig:p2_diagnostics}. 
The zero-energy momentum cuts in Figure~\ref{fig:p2_diagnostics}(\subref{fig:kxA1-p2}) show that the Fermi-level peak is strongly suppressed as $p_2$ increases. 
This trend is also reflected in Figure~\ref{fig:p2_diagnostics}(\subref{fig:A_p2}), where the extracted zero-energy peak height $A_{\rm peak}$ decreases with $p_2$. 
At the same time, the frequency cuts $A(\omega,k_F(p_2))$ in Figure~\ref{fig:p2_diagnostics}(\subref{fig:Aw-p2}) show that the dominant spectral peak moves to a finite positive frequency. 
The peak position $\omega_{\rm peak}$ in Figure~\ref{fig:p2_diagnostics}(\subref{fig:w-p2})  further captures this shift.

Taken together, these results show that the $p_2$ coupling suppresses the zero-energy response, transfers the dominant spectral weight to finite frequency, and produces a flat-bottom feature in the upper spectral branch. 
This resembles the flat-band-like spectra obtained in the holographic Ginzburg--Landau--Wilson framework of Ref.~\cite{Oh:2020cym}, but the origin is different. 
In Ref.~\cite{Oh:2020cym}, such structures arise from condensate order parameters dual to fermion bilinears, whereas in our setup the tensor channel is embedded in the two-current Gubser-Rocha background and is modulated by $F_{\alpha\beta}Y^{\alpha\beta}$ together with the axion sector. 
Thus, the spectral behavior driven by $p_2$ represents a background-modulated flat-bottom reconstruction, distinct from the fully developed hard gaps or flat-band phases.

We next consider the vector/axion-gradient coupling $p_3$. 
This interaction combines the current-mixing factor $F_{\alpha\beta}Y^{\alpha\beta}$, the axion gradient $\partial_\mu\psi_I$, and the chiral matrix $\Gamma^5$. 
It therefore acts as an axial-type deformation of the probe fermion sector, with an effective strength controlled by both the two-current background and the finite axion background.

The bottom row of Figure~\ref{fig:full_comparison} shows the spectral evolution for $p_3=0.2,0.5,0.8$. 
The zero-frequency density plots in Figures~\ref{fig:full_comparison}(\subref{fig:row2_1}), \ref{fig:full_comparison}(\subref{fig:row2_3}), and \ref{fig:full_comparison}(\subref{fig:row2_5}) show that increasing $p_3$ drives the spectral weight away from a closed Fermi ring and concentrates it on selected momentum segments. 
At larger $p_3$, this produces arc-like open structures. 
The corresponding dispersion plots in Figures~\ref{fig:full_comparison}(\subref{fig:row2_2}), \ref{fig:full_comparison}(\subref{fig:row2_4}), and \ref{fig:full_comparison}(\subref{fig:row2_6}) show an imbalance between the positive-momentum and negative-momentum branches. 
Thus, the main effect of the $p_3$ channel is a momentum-space redistribution of spectral weight, in contrast to the finite-frequency transfer observed for the $p_2$ channel.

This imbalance is further illustrated by the zero-frequency momentum cuts in Figure~\ref{p3-eta-akx-w}(\subref{fig:p3_kx_cuts}). 
As $p_3$ increases, the spectral weight on the negative-$k_x$ side is suppressed, while the dominant peak becomes concentrated on the positive-$k_x$ branch. 
To characterize this left-right imbalance, we define
\begin{equation}
W_+(p_3)=\int_{0}^{k_{\max}} A(\omega=0,k_x;p_3)\,dk_x,
\qquad
W_-(p_3)=\int_{-k_{\max}}^{0} A(\omega=0,k_x;p_3)\,dk_x .
\end{equation}
The normalized imbalance is then
\begin{equation}
R_W(p_3)=
\frac{W_+(p_3)-W_-(p_3)}
{W_+(p_3)+W_-(p_3)} .
\end{equation}
Here $k_{\max}$ is fixed for all values of $p_3$ and covers the momentum window shown in Figure~\ref{p3-eta-akx-w}(\subref{fig:p3_kx_cuts}). 
A value $R_W=0$ corresponds to a left-right symmetric distribution, while $R_W>0$ indicates that the spectral weight is biased toward the positive-$k_x$ branch. 
We use $R_W$ only as a diagnostic measure of the imbalance in $A(k_x)|_{\omega=0}$, not as an order parameter.

Figure~\ref{p3-eta-akx-w}(\subref{fig:p3_RW}) shows that $R_W$
increases rapidly with $p_3$ and approaches unity at larger coupling,
confirming an increasingly one-sided redistribution of the
zero-frequency spectral weight. This asymmetry originates from the
$p_3$ interaction, which combines the current-mixing factor
$F_{\alpha\beta}Y^{\alpha\beta}$ with the directional axion-gradient
structure $(\Gamma^\mu\partial_\mu\psi_I)\Gamma^5$. Although the
two-axion background remains homogeneous, this non-minimal coupling
induces a directional vector/axial deformation in the probe-fermion
sector and produces an imbalance between the positive- and
negative-momentum branches. The resulting open spectral segments are
reminiscent of Fermi-arc-like spectra generated by axial-vector-type
couplings~\cite{Oh:2020cym}, but their origin is distinct: here no
independent condensate order parameter is introduced, and the
reconstruction is driven directly by the current-mixing-modulated
axion-dependent interaction.

\subsection{Dependence on Dissipation and Doping}

\begin{figure}[htbp]
    \centering
    \captionsetup[subfigure]{labelformat=parens}

    \begin{subfigure}[b]{0.46\linewidth}
        \centering
        \includegraphics[width=\linewidth]{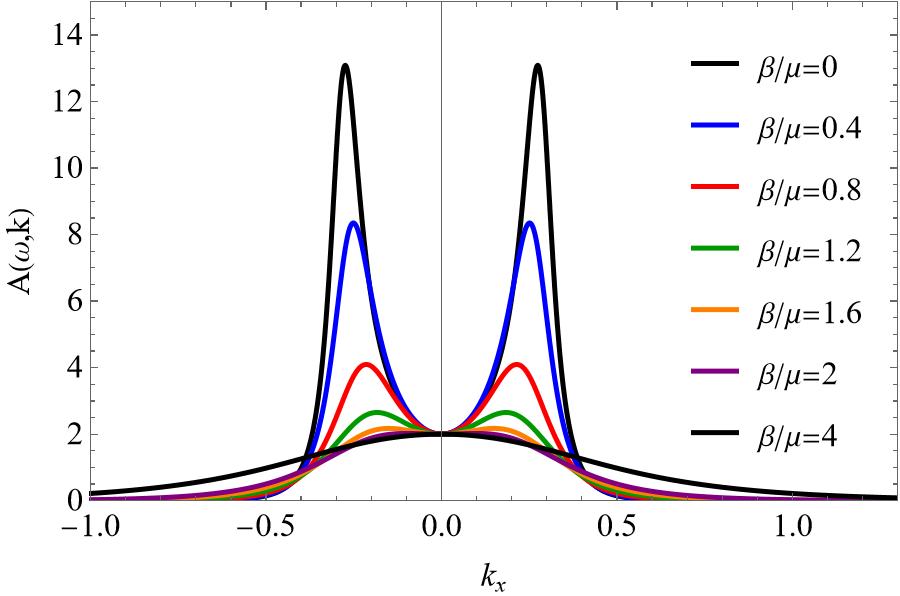}
        \caption{\centering $\chi=1$ (Varying \(\beta/\mu\))}
        \label{fig:beta_chi1}
    \end{subfigure}
    \hspace{0.03\linewidth}
    \begin{subfigure}[b]{0.46\linewidth}
        \centering
        \includegraphics[width=\linewidth]{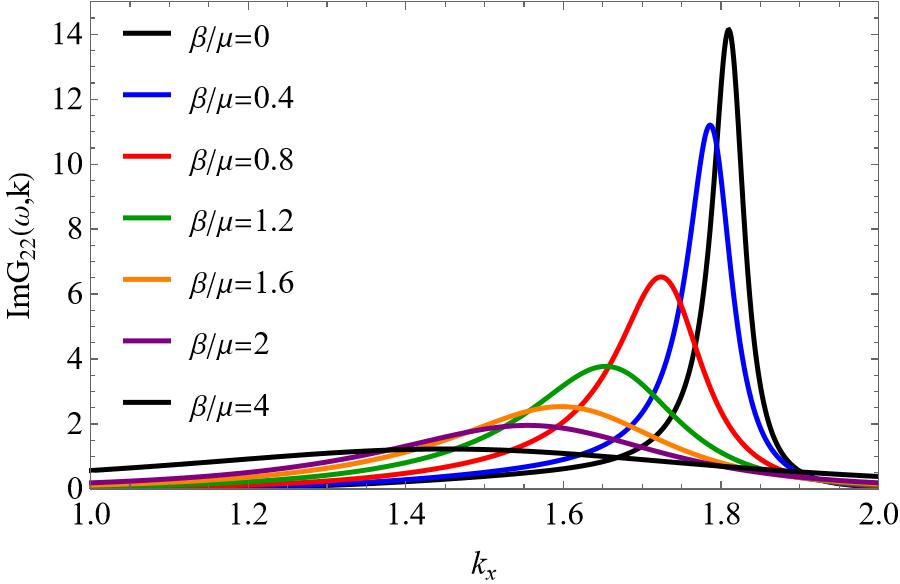}
        \caption{\centering $p_1=4$ (Varying \(\beta/\mu\))}
        \label{fig:beta_p1}
    \end{subfigure}

    \vspace{0.6em}

    \begin{subfigure}[b]{0.46\linewidth}
        \centering
        \includegraphics[width=\linewidth]{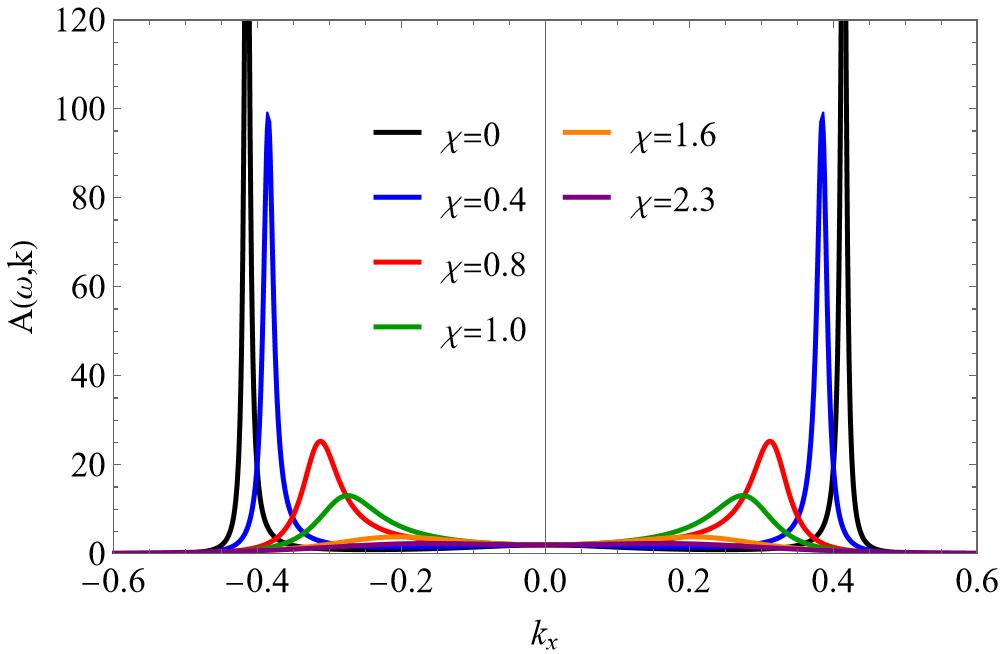}
        \caption{\centering $\beta/\mu=0$ (Varying $\chi$)}
        \label{fig:chi_beta0}
    \end{subfigure}
    \hspace{0.03\linewidth}
    \begin{subfigure}[b]{0.46\linewidth}
        \centering
        \includegraphics[width=\linewidth]{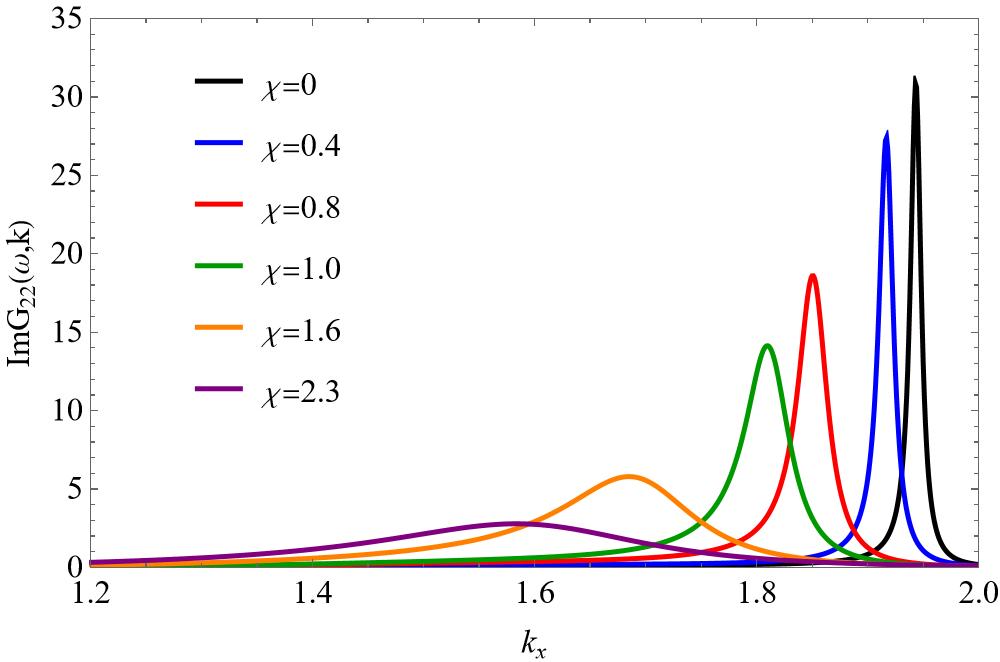}
        \caption{\centering $p_1=4$ (Varying $\chi$)}
        \label{fig:chi_p1}
    \end{subfigure}

    \caption{
    Zero-frequency momentum cuts under varying background parameters. 
The left column shows the baseline spectral function \(A(k_x)\), while the 
right column shows the dipole-channel component \(\mathrm{Im}G_{22}(k_x)\) 
at \(p_1=4\). Top row: varying \(\beta/\mu\) at fixed $\chi=1.0$. Bottom row: varying \(\chi\) at fixed $\beta/\mu=0$. 
Other fixed parameters are \(m_\psi=0\), \(q=1\), and \(T/\mu=0.009\).
    }
    \label{fig:stability_analysis_beta_chi}
\end{figure}

Having established the spectral reconstructions induced by the non-minimal couplings, we now examine their dependence on the background parameters.  
In the two-current Gubser-Rocha geometry, both the momentum-relaxation strength $\beta/\mu$ and the doping parameter $\chi$ modify the background fields, thereby tuning the fermionic response. 
Figure~\ref{fig:stability_analysis_beta_chi} compares the zero-frequency
total spectral function $A(k_x)$ for the baseline case with the
dipole-channel component $\mathrm{Im}\,G_{22}(k_x)$ at $p_1=4$ under
variations of $\beta/\mu$ and $\chi$.

For the baseline spectra without additional non-minimal couplings, Figure~\ref{fig:stability_analysis_beta_chi}(\subref{fig:beta_chi1}) shows that increasing $\beta/\mu$ broadens the two Fermi peaks and reduces their height. 
At strong momentum relaxation, these peaks become poorly resolved, and the spectral weight is distributed over a wider momentum range. 
This reflects the loss of coherence caused by momentum relaxation. 
By contrast, as seen in Figure~\ref{fig:stability_analysis_beta_chi}(\subref{fig:chi_beta0}), increasing $\chi$ primarily shifts the peaks toward smaller momenta and attenuates their amplitudes. 
This indicates a contraction and weakening of the underlying Fermi surface.

The right column shows the corresponding response in the dipole channel at fixed $p_1=4$. 
When varying $\beta/\mu$, Figure~\ref{fig:stability_analysis_beta_chi}(\subref{fig:beta_p1}) shows a pronounced reduction and broadening of the dominant finite-momentum peak. 
Although the peak position also moves toward smaller momentum, the most visible effect of increasing $\beta/\mu$ is the loss of spectral coherence. 
For varying $\chi$, Figure~\ref{fig:stability_analysis_beta_chi}(\subref{fig:chi_p1}) shows a different tendency: the dominant peak shifts continuously to smaller momentum as $\chi$ increases, while the peak remains relatively well defined over a broad range of doping. 

Thus, $\beta/\mu$ and $\chi$ tune the dipole-channel spectrum in different ways. 
Momentum relaxation primarily broadens and weakens the peak, whereas doping mainly shifts the Fermi momentum and reduces the spectral intensity. 
Although the finite-momentum peak survives within the parameter regime considered here, its position, width, and amplitude are renormalized   differently by these two background parameters.

\begin{figure}[htbp]
    \centering
    \captionsetup[subfigure]{labelformat=parens, labelsep=space}

    \begin{subfigure}[b]{0.46\linewidth}
        \centering
        \includegraphics[width=\linewidth]{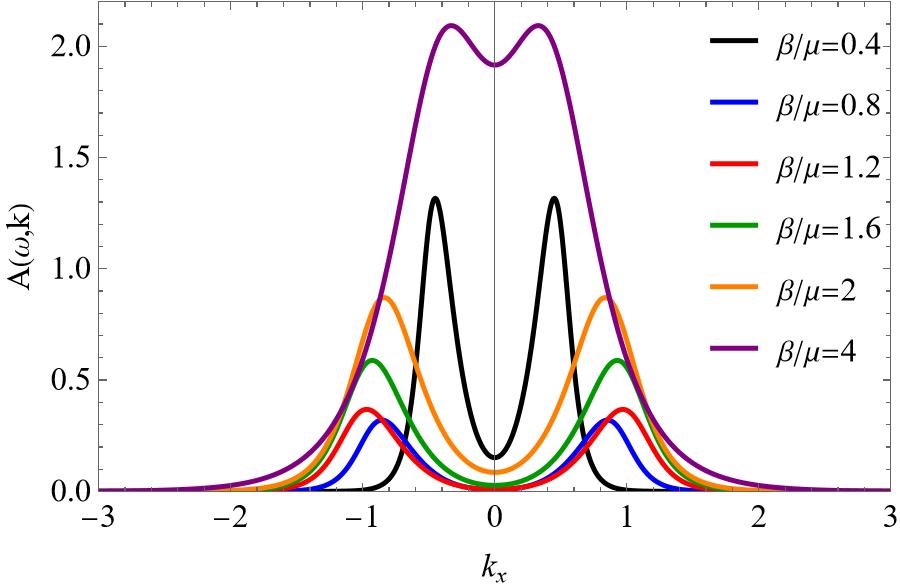}
        \caption{\centering $p_2=0.4$ (Varying \(\beta/\mu\))}
        \label{fig:p2_beta_kx}
    \end{subfigure}
    \hspace{0.03\linewidth}
    \begin{subfigure}[b]{0.46\linewidth}
        \centering
        \includegraphics[width=\linewidth]{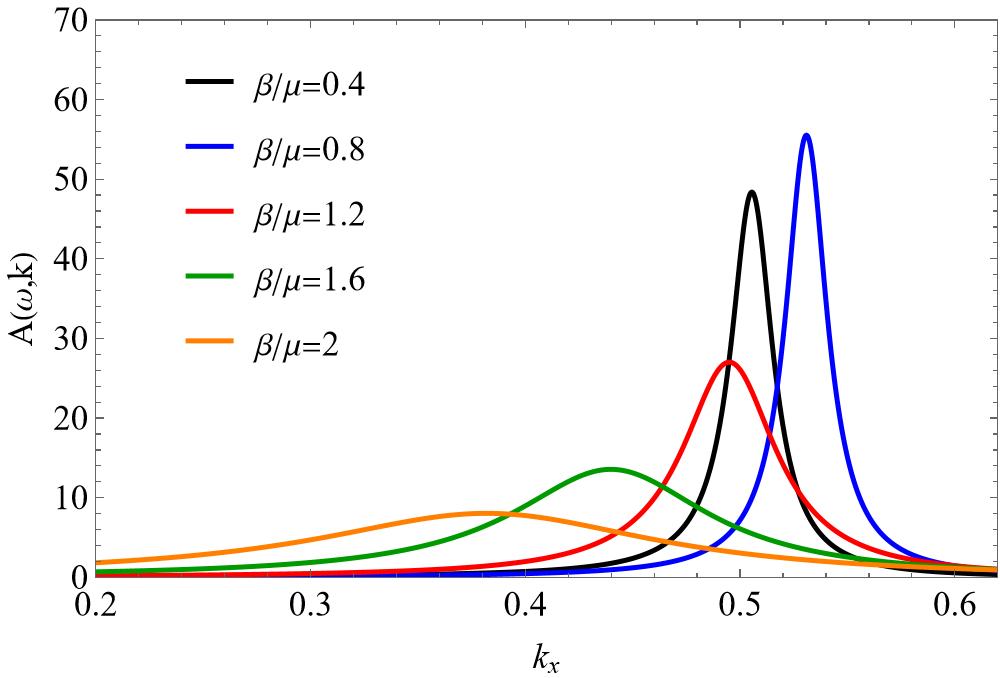}
        \caption{\centering $p_3=0.5$ (Varying \(\beta/\mu\))}
        \label{fig:p3_beta_kx}
    \end{subfigure}

    \vspace{0.6em}

    \begin{subfigure}[b]{0.46\linewidth}
        \centering
        \includegraphics[width=\linewidth]{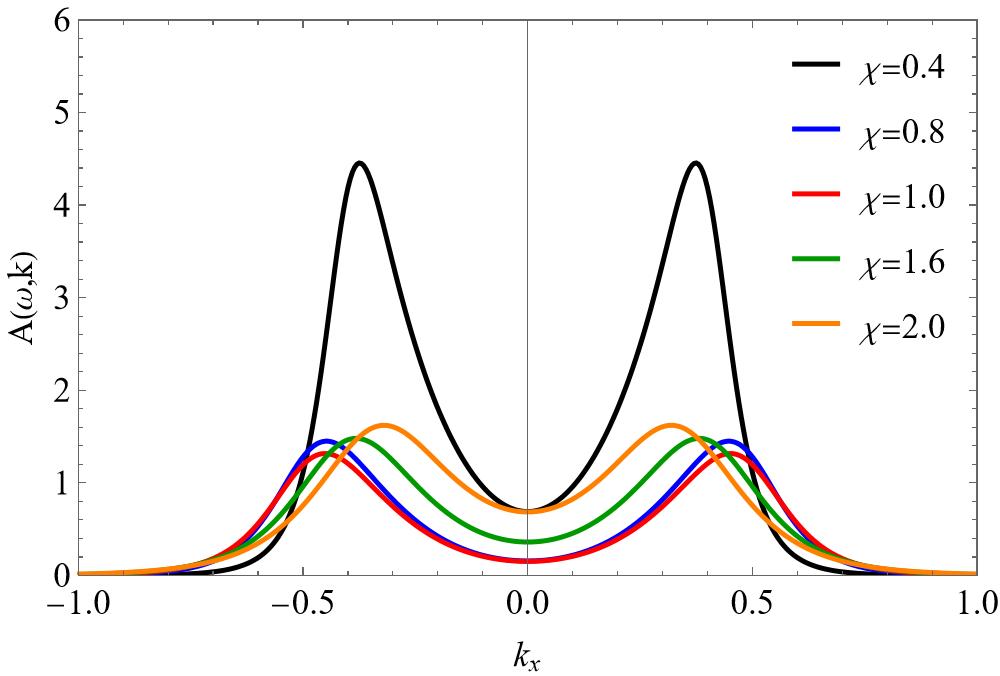}
        \caption{\centering $p_2=0.4$ (Varying $\chi$)}
        \label{fig:p2_chi_kx}
    \end{subfigure}
    \hspace{0.03\linewidth}
    \begin{subfigure}[b]{0.46\linewidth}
        \centering
        \includegraphics[width=\linewidth]{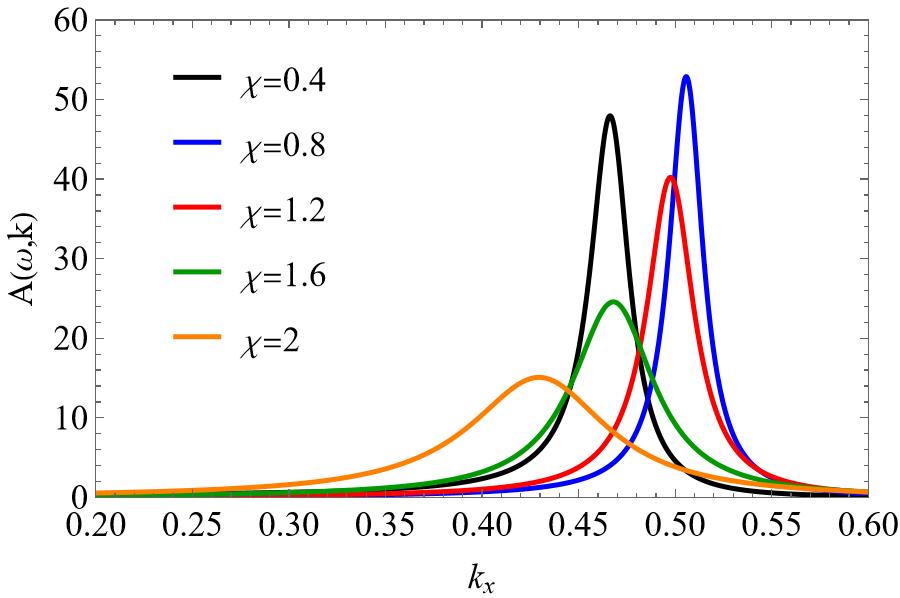}
        \caption{\centering $p_3=0.5$ (Varying $\chi$)}
        \label{fig:p3_chi_kx}
    \end{subfigure}

    \caption{
    Spectral function cuts $A(k_x)$ at $\omega=0$ under varying background parameters.
    Top row: varying $\beta/\mu$ at fixed $\chi=1.0$.
    Bottom row: varying $\chi$ at fixed $\beta/\mu=0.4$.
    Columns correspond to the tensor channel $p_2=0.4$ (left) and the vector/axion-gradient channel $p_3=0.5$ (right).
    Other fixed parameters are $m_\psi=0$, $q=1$, and $T/\mu=0.009$.
    }
    \label{fig:stability_analysis_p3p2}
\end{figure}

The dependence on $\beta/\mu$ and $\chi$ becomes more involved in the
axion-related $p_2$ and $p_3$ channels. Unlike the dipole coupling,
these interactions depend explicitly on the axion sector and are also
modulated by the current-mixing factor
$F_{\alpha\beta}Y^{\alpha\beta}$. Therefore, varying $\beta/\mu$ or
$\chi$ changes not only the background geometry but also the effective
strength of the corresponding non-minimal interactions.
Figure~\ref{fig:stability_analysis_p3p2} shows the resulting
zero-frequency momentum cuts for $p_2=0.4$ and $p_3=0.5$.

For the tensor channel, Figure~\ref{fig:stability_analysis_p3p2}(\subref{fig:p2_beta_kx}) shows that increasing $\beta/\mu$ broadens the symmetric two-peak structure and gradually fills the central low-spectral-weight region. 
This indicates that momentum relaxation weakens the contrast of the low-energy suppression induced by $p_2$. 
In Figure~\ref{fig:stability_analysis_p3p2}(\subref{fig:p2_chi_kx}), increasing $\chi$ also reduces the peak height and broadens the distribution, while the off-center two-peak profile remains visible. 
Thus, the $p_2$ response is mainly affected through the broadening and weakening of its symmetric low-energy structure.

Turning to the vector/axion-gradient channel, the response exhibits a distinct behavior. 
Figures~\ref{fig:stability_analysis_p3p2}(\subref{fig:p3_beta_kx}) and \ref{fig:stability_analysis_p3p2}(\subref{fig:p3_chi_kx})  reveal that the positive-momentum peak evolves non-monotonically under both $\beta/\mu$ and $\chi$: it is enhanced at intermediate values, but becomes broader, weaker, and shifts toward smaller momentum at larger values. 
This suggests that the $p_3$ channel is controlled by the combined effect of the axion sector, current mixing, and dissipative broadening. 
Therefore, while $p_2$ retains a symmetric two-peak structure whose contrast is reduced by the background, $p_3$ produces a one-sided spectral feature that is tuned more sensitively and non-monotonically by the dissipative two-current background.

In summary, the background-parameter scan shows that $\beta/\mu$ and $\chi$ play different roles in the reconstructed spectra. 
Momentum relaxation mainly broadens the spectral peaks and reduces their coherence, while doping primarily changes the charge-sector background and shifts the characteristic momentum scale. 
For the $p_2$ channel, these effects weaken the contrast of the symmetric low-energy suppression. 
For the $p_3$ channel, they produce a non-monotonic tuning of the one-sided spectral peak. 
This confirms that the $p_2$ and $p_3$ reconstructions are not determined by the bare couplings alone, but are controlled by the dissipative two-current Gubser-Rocha background.
\section{Summary}
\label{summary}

In this work, we investigated fermionic spectral functions in a two-current Gubser-Rocha background with linear-axion momentum relaxation. 
The main purpose was to characterize how non-minimal fermion couplings reconstruct the spectral response in a dissipative two-current geometry. 
We first analyzed the standard dipole coupling $p_1$ as the conventional Pauli-type channel in this background. 
We then introduced the tensor coupling $p_2$ and the vector/axion-gradient coupling $p_3$ as current-mixing-modulated axion-related channels.

For the dipole coupling $p_1$, the zero-frequency spectral weight remains distributed along a closed Fermi surface, while the Fermi momentum shifts outward as $p_1$ increases. The density of states develops a particle--hole-asymmetric profile, and $\mathrm{DOS}(0)$ shows a non-monotonic dependence on $p_1$ but remains finite over the parameter range considered. This non-monotonic behavior reflects a competition between the outward shift of the Fermi surface, which initially increases the momentum-space phase-space contribution, and the suppression of low-energy spectral weight at stronger coupling. Consequently, no clean hard Mott gap is resolved in the present two-current Gubser--Rocha background. Instead, the dipole coupling produces a soft, particle--hole-asymmetric low-energy reconstruction reminiscent of the Mott-like behavior reported in holographic fermion systems.

The tensor coupling $p_2$ yields a distinct type of spectral reconstruction. This channel suppresses the zero-energy response and shifts the dominant spectral weight to finite frequencies, while the upper spectral branch develops a pronounced flat-bottom feature. At the level of the bulk Dirac equation, this behavior is associated with the tensor structure $\Gamma^{\mu\nu}H_{\mu\nu}$, which modifies the bulk fermion dynamics and redistributes spectral weight away from the Fermi level. Thus, the main effect of $p_2$ is low-energy suppression accompanied by spectral-weight transfer to finite frequencies, rather than the formation of a fully developed hard gap or a genuine flat-band phase. This flattening tendency resembles the flat-band-like spectra found in holographic Ginzburg--Landau--Wilson constructions, but its physical origin is different because no independent condensate order parameter is introduced. Instead, the interaction involves both the auxiliary tensor $H_{\mu\nu}$ constructed from the axion gradients and the current-mixing factor $F_{\alpha\beta}Y^{\alpha\beta}$.

For the vector/axion-gradient coupling $p_3$, the spectral response is characterized primarily by an asymmetric redistribution of spectral weight in momentum space. The zero-frequency spectral weight becomes concentrated on selected momentum segments, producing arc-like open structures. This asymmetry is further quantified by the integrated imbalance ratio $R_W$, which shows that the spectral weight becomes increasingly biased toward the positive-momentum branch as $p_3$ increases. The origin of this behavior lies in the axial-vector structure $(\Gamma^\mu\partial_\mu\psi_I)\Gamma^5$, which affects the two momentum branches differently and produces a directional redistribution of spectral weight even though the axion background itself remains homogeneous. The distinctive feature of the present construction is that the strength of this deformation is controlled jointly by the axion gradient and the mixed field-strength contraction.

We also examined how the reconstructed spectra respond to the background parameters $\beta/\mu$ and $\chi$. 
These two parameters affect the spectra in different ways. 
Momentum relaxation mainly broadens the spectral peaks and reduces their coherence, while the doping parameter primarily shifts the characteristic momentum scale through the charge-sector background. 
For the dipole channel, these effects continuously renormalize the finite-momentum peak. 
For the $p_2$ channel, they weaken the contrast of the symmetric low-energy suppression. 
For the $p_3$ channel, they tune the one-sided spectral segment in a non-monotonic manner. 
These results show that the spectral reconstructions induced by the non-minimal couplings are not determined solely by the bare coupling constants, but are also sensitive to the dissipative two-current background.

Building on these results, a natural next step is to study the interplay between the $p_2$ and $p_3$ interactions when both are present and to examine the resulting spectral regimes under variations of $T/\mu$, the fermion charge $q$, the doping parameter $\chi$, and the momentum-relaxation scale $\beta/\mu$. Introducing a charged scalar sector would further allow one to explore how the reconstructed normal-state spectra influence the superconducting instability and the gap structure in the condensed phase. More general momentum-relaxing backgrounds, together with transport and two-particle response calculations, could further connect the single-particle spectral reconstruction found here with collective phenomena such as superconductivity and density-wave order.

\section*{Acknowledgments}
We thank Hai-cang Ren and  Yan-Qing Zhao for useful discussions. This work is supported in part by the National Key Research and Development Program of China under Contract No. 2022YFA1604900. This work is also partly supported by the National Natural Science Foundation of China(NSFC) under Grants No. 12435009, and No. 12275104.

\appendix

\section{Radial Flow Equations for Axion-Induced Interactions}
\label{app:flow_equations}

In this appendix, we provide the detailed derivation of the radial evolution equations for the fermionic sectors coupled to the axion background. Specifically, we focus on the tensor ($p_2$) and vector ($p_3$) channels introduced in Sec.~\ref{subsec:Interaction_Driven}. These derivations underpin the numerical results discussed in the main text.

\subsection{Tensor Coupling ($p_2$)}

We begin with the tensor coupling channel defined by the interaction term $\mathcal{L}^{\text{(II-a)}}_{\text{int}}$, where the bulk fermions couple to the auxiliary tensor field $H_{\mu\nu}$. The corresponding Dirac equation is:
\begin{equation}
    \left(\Gamma^a D_a - m_\psi - ip_2 F_{\alpha\beta}Y^{\alpha\beta}\Gamma^{\mu\nu}H_{\mu\nu}\right) \psi = 0 \,.
\end{equation}
Upon substituting the explicit forms of the background fields and the metric ansatz, this equation expands to:
\begin{align}
   &(\Gamma^t\sqrt{g^{tt}}(\partial_t-iq A_t)+\Gamma^r\sqrt{g^{rr}}\partial_r+\Gamma^x\sqrt{g^{xx}}\partial_x+\Gamma^y\sqrt{g^{yy}}\partial_y)\psi \nonumber \\
   &\qquad+\sqrt{g^{rr}}\partial_r\ln{(-g_{tt}g_{ii}})^{\frac{1}{4}}\Gamma^r\psi-4ip_2g^{rr}g^{tt}A_t' B_t'\sqrt{g^{xx}g^{yy}}\Gamma^{xy}H_{xy}\psi
   -m_\psi \psi=0 \,.
\end{align}
To solve this system numerically, we perform a Fourier transformation and project the 4-component spinor onto two 2-component spinors, $\phi = (\phi_1, \phi_2)^T$. The resulting coupled radial equations are:
\begin{align}\label{A3}
\frac{\sqrt{g_{ii}}}{\sqrt{g_{rr}}}\partial_r \otimes \begin{pmatrix}\phi_1 \\ \phi_2 \end{pmatrix} 
&- i\frac{\sqrt{g_{ii}}}{\sqrt{g_{tt}}}(\omega + qA_t) \sigma_2 \otimes \begin{pmatrix} \phi_1 \\ \phi_2 \end{pmatrix} 
+ m_\psi\sqrt{g_{ii}}\sigma_3 \otimes \begin{pmatrix} \phi_1 \\ \phi_2 \end{pmatrix} 
\pm k_x\sigma_1 \otimes \begin{pmatrix} \phi_1 \\ \phi_2 \end{pmatrix} \nonumber \\
&
- k_y\sigma_1 \otimes \begin{pmatrix} \phi_2 \\ \phi_1 \end{pmatrix}  \mp \frac{2 i p_2 \beta^2  A_t' B_t'}{g_{rr} g_{tt}\sqrt{g_{ii}}} \sigma_3 \otimes \begin{pmatrix} \phi_2 \\ \phi_1 \end{pmatrix} 
= 0 \,.
\end{align}
Further decomposing the spinors into scalar components $\phi_I = (y_I, z_I)^T$ (for $I=1,2$) yields the following set of first-order differential equations:
\begin{align}
&\frac{\sqrt{g_{ii}} }{\sqrt{g_{rr}}}\otimes\begin{pmatrix} y_1' \\ z_1' \end{pmatrix} \pm m_\psi \sqrt{g_{ii}}\otimes \begin{pmatrix} y_1 \\ z_1 \end{pmatrix}\mp\frac{\sqrt{g_{ii}}}{\sqrt{g_{tt}}}(\omega+q A_t)\otimes \begin{pmatrix} z_1 \\ y_1 \end{pmatrix}+ k_x\begin{pmatrix} z_1 \\ y_1 \end{pmatrix} \nonumber \\
&\qquad -k_y\otimes \begin{pmatrix} z_2 \\ y_2 \end{pmatrix} \mp \frac{2 i p_2 \beta^2  A_t' B_t'}{g_{rr} g_{tt}\sqrt{g_{ii}}} \otimes \begin{pmatrix} y_2 \\ z_2 \end{pmatrix}=0 \\
&\frac{\sqrt{g_{ii}} }{\sqrt{g_{rr}}}\otimes\begin{pmatrix} y_2' \\ z_2' \end{pmatrix} \pm m_\psi \sqrt{g_{ii}}\otimes \begin{pmatrix} y_2 \\ z_2 \end{pmatrix}\mp\frac{\sqrt{g_{ii}}}{\sqrt{g_{tt}}}(\omega+q A_t)\otimes \begin{pmatrix} z_2 \\ y_2 \end{pmatrix}- k_x\begin{pmatrix} z_2 \\ y_2 \end{pmatrix} \nonumber \\
&\qquad -k_y\otimes \begin{pmatrix} z_1 \\ y_1 \end{pmatrix} \pm \frac{2 i p_2 \beta^2  A_t' B_t'}{g_{rr} g_{tt}\sqrt{g_{ii}}} \otimes \begin{pmatrix} y_1 \\ z_1 \end{pmatrix}=0 \,.
\end{align}

Finally, employing the definition of the Green's function matrix $\tilde{G}$, we obtain the radial flow equations required for our numerical analysis:
\begin{align}
    \frac{\tilde{G}_{11}'}{\sqrt{g_{rr}}} + 2m_\psi \tilde{G}_{11} &= -\frac{k_{x}}{\sqrt{g_{xx}}}(1+\tilde{G}_{12}\tilde{G}_{21}-\tilde{G}_{11}^2)-\frac{k_y \tilde{G}_{11}(\tilde{G}_{12}+\tilde{G}_{21})}{\sqrt{g_{yy}}}\nonumber \\
    &\quad +\frac{(\omega+q A_t)(1+\tilde{G}_{11}^2+\tilde{G}_{12}\tilde{G}_{21})}{\sqrt{g_{tt}}} - \frac{2 i p_2 \beta^2 A_t' B_t' (\tilde{G}_{12} - \tilde{G}_{21})}{g_{xx}g_{rr} g_{tt}} \\
    \frac{\tilde{G}_{12}'}{\sqrt{g_{rr}}} + 2m_\psi \tilde{G}_{12} &= \frac{k_{y}}{\sqrt{g_{yy}}}(1-\tilde{G}_{12}^2-\tilde{G}_{11}\tilde{G}_{22})+\frac{k_x \tilde{G}_{12}(\tilde{G}_{11}-\tilde{G}_{22})}{\sqrt{g_{xx}}}\nonumber \\ 
    &\quad +\frac{(\omega+q A_t)(\tilde{G}_{11}\tilde{G}_{12}+\tilde{G}_{12}\tilde{G}_{22})}{\sqrt{g_{tt}}} + \frac{2 i p_2 \beta^2 A_t' B_t' (\tilde{G}_{11} + \tilde{G}_{22})}{g_{xx}g_{rr} g_{tt}}
\end{align}
    \begin{align}    
    \frac{\tilde{G}_{21}'}{\sqrt{g_{rr}}} + 2m_\psi \tilde{G}_{21} &= \frac{k_{y}}{\sqrt{g_{yy}}}(1-\tilde{G}_{21}^2-\tilde{G}_{11}\tilde{G}_{22})+\frac{k_x \tilde{G}_{21}(\tilde{G}_{11}-\tilde{G}_{22})}{\sqrt{g_{xx}}}\nonumber \\ 
    &\quad +\frac{(\omega+q A_t)(\tilde{G}_{11}\tilde{G}_{21}+\tilde{G}_{21}\tilde{G}_{22})}{\sqrt{g_{tt}}} - \frac{2 i p_2 \beta^2 A_t' B_t' (\tilde{G}_{11} + \tilde{G}_{22})}{g_{xx}g_{rr} g_{tt}}\\
    \frac{\tilde{G}_{22}'}{\sqrt{g_{rr}}} + 2m_\psi \tilde{G}_{22} &= \frac{k_{x}}{\sqrt{g_{xx}}}(1+\tilde{G}_{12}\tilde{G}_{21}-\tilde{G}_{22}^2)-\frac{k_y \tilde{G}_{22}(\tilde{G}_{12}+\tilde{G}_{21})}{\sqrt{g_{yy}}}\nonumber \\ 
    &\quad +\frac{(\omega+q A_t)(1+\tilde{G}_{22}^2+\tilde{G}_{12}\tilde{G}_{21})}{\sqrt{g_{tt}}} - \frac{2 i p_2 \beta^2 A_t' B_t' (\tilde{G}_{12} - \tilde{G}_{21})}{g_{xx}g_{rr} g_{tt}} \,.
\end{align}

\subsection{Vector Coupling ($p_3$)}

Next, we consider the vector coupling channel $\mathcal{L}^{\text{(II-b)}}_{\text{int}}$, which couples the axion gradient to the axial current. The modified Dirac equation reads:
\begin{equation}
    \left(\Gamma^a D_a - m_\psi - ip_3F_{\alpha\beta}Y^{\alpha\beta}\sum_{I=1}^{2} (\Gamma^{\mu} \partial_{\mu}\psi_I) \Gamma^5\right) \psi = 0 \,.
\end{equation}
Substituting the relevant background fields into the interaction term, the equation expands as:
\begin{align}
   &(\Gamma^t\sqrt{g^{tt}}(\partial_t-iq A_t)+\Gamma^r\sqrt{g^{rr}}\partial_r+\Gamma^x\sqrt{g^{xx}}\partial_x+\Gamma^y\sqrt{g^{yy}}\partial_y)\psi \nonumber \\
   &\qquad+\sqrt{g^{rr}}\partial_r\ln{(-g_{tt}g_{ii}})^{\frac{1}{4}}\Gamma^r\psi-2ip_3g^{rr}g^{tt}\sqrt{g^{ii}}A_t' B_t'(\Gamma^{x}\partial_x\psi_I+\Gamma^{y}\partial_y\psi_I)\Gamma^5\psi
   -m_\psi \psi=0 \,\end{align}
Following the  spinor reduction procedure, the equation of motion for the 2-component spinor fields takes the form:
\begin{align}
\frac{\sqrt{g_{ii}}}{\sqrt{g_{rr}}}\partial_r \otimes \begin{pmatrix}\phi_1 \\ \phi_2 \end{pmatrix} 
&- i\frac{\sqrt{g_{ii}}}{\sqrt{g_{tt}}}(\omega + qA_t) \sigma_2 \otimes \begin{pmatrix} \phi_1 \\ \phi_2 \end{pmatrix} 
+ m_\psi\sqrt{g_{ii}}\sigma_3 \otimes \begin{pmatrix} \phi_1 \\ \phi_2 \end{pmatrix} 
\pm k_x\sigma_1 \otimes \begin{pmatrix} \phi_1 \\ \phi_2 \end{pmatrix} \nonumber \\
&
- k_y\sigma_1 \otimes \begin{pmatrix} \phi_2 \\ \phi_1 \end{pmatrix} + \frac{2 p_3 \beta  A_t' B_t'}{g_{rr} g_{tt}} \sigma_3 \otimes \begin{pmatrix} \phi_1 + \phi_2 \\ \phi_1 - \phi_2 \end{pmatrix} 
= 0 \,.
\end{align}
Rewriting this in terms of the scalar components $y_I$ and $z_I$ leads to the coupled system:
\begin{align}
&\frac{\sqrt{g_{ii}} }{\sqrt{g_{rr}}}\otimes\begin{pmatrix} y_1' \\ z_1' \end{pmatrix} \pm m_\psi \sqrt{g_{ii}}\otimes \begin{pmatrix} y_1 \\ z_1 \end{pmatrix}\mp\frac{\sqrt{g_{ii}}}{\sqrt{g_{tt}}}(\omega+q A_t)\otimes \begin{pmatrix} z_1 \\ y_1 \end{pmatrix}+ k_x\begin{pmatrix} z_1 \\ y_1 \end{pmatrix} \nonumber \\
&\qquad -k_y\otimes \begin{pmatrix} z_2 \\ y_2 \end{pmatrix} 
+ \frac{2 p_3 \beta  A_t' B_t'}{g_{rr} g_{tt}} \otimes \begin{pmatrix} y_1 + y_2 \\ -(z_1 + z_2) \end{pmatrix}=0 \\
&\frac{\sqrt{g_{ii}} }{\sqrt{g_{rr}}}\otimes\begin{pmatrix} y_2' \\ z_2' \end{pmatrix} \pm m_\psi \sqrt{g_{ii}}\otimes \begin{pmatrix} y_2 \\ z_2 \end{pmatrix}\mp\frac{\sqrt{g_{ii}}}{\sqrt{g_{tt}}}(\omega+q A_t)\otimes \begin{pmatrix} z_2 \\ y_2 \end{pmatrix}- k_x\begin{pmatrix} z_2 \\ y_2 \end{pmatrix} \nonumber \\
&\qquad -k_y\otimes \begin{pmatrix} z_1 \\ y_1 \end{pmatrix} 
+ \frac{2 p_3 \beta  A_t' B_t'}{g_{rr} g_{tt}} \otimes \begin{pmatrix} y_1 - y_2 \\ -(z_1 - z_2) \end{pmatrix}=0 \,.
\end{align}
The resulting flow equations, which are used to compute the $p_3$-induced asymmetric spectral redistribution, are given by:
\begin{align}
    \frac{\tilde{G}_{11}'}{\sqrt{g_{rr}}} + 2m_\psi \tilde{G}_{11} &= -\frac{k_{x}}{\sqrt{g_{xx}}}(1+\tilde{G}_{12}\tilde{G}_{21}-\tilde{G}_{11}^2)-\frac{k_y \tilde{G}_{11}(\tilde{G}_{12}+\tilde{G}_{21})}{\sqrt{g_{yy}}}\nonumber \\
    &\quad +\frac{(\omega+q A_t)(1+\tilde{G}_{11}^2+\tilde{G}_{12}\tilde{G}_{21})}{\sqrt{g_{tt}}} - \frac{2 p_3 \beta A_t' B_t'}{g_{rr} g_{tt}\sqrt{g_{ii}}} (2\tilde{G}_{11} + \tilde{G}_{12} + \tilde{G}_{21}) \\
    \frac{\tilde{G}_{12}'}{\sqrt{g_{rr}}} + 2m_\psi \tilde{G}_{12} &= \frac{k_{y}}{\sqrt{g_{yy}}}(1-\tilde{G}_{12}^2-\tilde{G}_{11}\tilde{G}_{22})+\frac{k_x \tilde{G}_{12}(\tilde{G}_{11}-\tilde{G}_{22})}{\sqrt{g_{xx}}}\nonumber \\ 
    &\quad +\frac{(\omega+q A_t)(\tilde{G}_{11}\tilde{G}_{12}+\tilde{G}_{12}\tilde{G}_{22})}{\sqrt{g_{tt}}} - \frac{2 p_3 \beta A_t' B_t'}{g_{rr} g_{tt}\sqrt{g_{ii}}} (\tilde{G}_{11} + \tilde{G}_{22})\\
    \frac{\tilde{G}_{21}'}{\sqrt{g_{rr}}} + 2m_\psi \tilde{G}_{21} &= \frac{k_{y}}{\sqrt{g_{yy}}}(1-\tilde{G}_{21}^2-\tilde{G}_{11}\tilde{G}_{22})+\frac{k_x \tilde{G}_{21}(\tilde{G}_{11}-\tilde{G}_{22})}{\sqrt{g_{xx}}}\nonumber \\ 
    &\quad +\frac{(\omega+q A_t)(\tilde{G}_{11}\tilde{G}_{21}+\tilde{G}_{21}\tilde{G}_{22})}{\sqrt{g_{tt}}} - \frac{2 p_3 \beta A_t' B_t'}{g_{rr} g_{tt}\sqrt{g_{ii}}} (\tilde{G}_{11} + \tilde{G}_{22})\\
    \frac{\tilde{G}_{22}'}{\sqrt{g_{rr}}} + 2m_\psi \tilde{G}_{22} &= \frac{k_{x}}{\sqrt{g_{xx}}}(1+\tilde{G}_{12}\tilde{G}_{21}-\tilde{G}_{22}^2)-\frac{k_y \tilde{G}_{22}(\tilde{G}_{12}+\tilde{G}_{21})}{\sqrt{g_{yy}}}\nonumber \\ 
    &\quad +\frac{(\omega+q A_t)(1+\tilde{G}_{22}^2+\tilde{G}_{12}\tilde{G}_{21})}{\sqrt{g_{tt}}} - \frac{2 p_3 \beta A_t' B_t'}{g_{rr} g_{tt}\sqrt{g_{ii}}} (\tilde{G}_{12} + \tilde{G}_{21} - 2\tilde{G}_{22})
\end{align}

\bibliographystyle{jhep}
\bibliography{refpapers.bib}
\end{document}